\documentclass[a4paper, 11pt]{article}
\RequirePackage[OT1]{fontenc}
\usepackage{amsfonts,amssymb,graphics,epsfig,verbatim,bm,latexsym,amsmath,url,float,amsbsy,multirow,multicol,rotating,enumerate,graphicx,pgf,tikz,subfig,color,xcolor,lscape,subfig}
\usepackage[textheight=722pt,textwidth=500pt,centering,headheight=50pt,headsep=12pt,footskip=12pt]{geometry}
\usepackage[colorlinks,citecolor=blue,linkcolor=purple!60!black,urlcolor=purple!60!black]{hyperref}
\usepackage[framemethod=TikZ]{mdframed}
\usepackage[numbers,sort&compress]{natbib}
\usetikzlibrary{positioning}
\usetikzlibrary[shapes.misc]
\usetikzlibrary{arrows,automata}
\allowdisplaybreaks
\makeatletter
\let\@fnsymbol\@arabic
\makeatother

\title{\vspace*{-2cm}\LARGE\bf Causal Inference for Continuous Multiple Time Point Interventions}
\date{}
\author{Michael Schomaker\footnote{Department of Statistics, Ludwig-Maximilians University, Munich, Germany, \href{mailto:michael.schomaker@stat.uni-muenchen.de}{michael.schomaker@stat.uni-muenchen.de}}~$^{,}$\footnote{Centre for Infectious Disease Epidemiology and Research, University of Cape Town, Cape Town, South Africa}~$^{,}$\footnote{Institute of Public Health, Medical Decision Making and Health Technology Assessment, UMIT -- University for Health Sciences, Medical Informatics and Technology, Hall in Tirol, Austria}
       \and
       Helen McIlleron\footnote{Division of Clinical Pharmacology, Department of Medicine, University of Cape Town, South Africa}~$^{,}$\footnote{Wellcome Centre for Infectious Diseases Research in Africa (CIDRI-Africa), Institute of Infectious Disease and Molecular Medicine, University of Cape Town, Cape Town, South Africa}
       \and
       Paolo Denti$^{\text{4}}$
       \and
       Iván Díaz\thanks{Division of Biostatistics, Department of Population Health, New York University Grossman School of Medicine; New York, United States of America}
       }

\begin{document}

\maketitle

\begin{abstract}
There are limited options to estimate the treatment effects of variables which are continuous and measured at multiple time points, particularly if the true dose-response curve should be estimated as closely as possible. However, these situations may be of relevance: in pharmacology, one may be interested in how outcomes of people living with -and treated for- HIV, such as viral failure, would vary for time-varying interventions such as different drug concentration trajectories. A challenge for doing causal inference with continuous interventions is that the positivity assumption is typically violated. To address positivity violations, we develop projection functions, which reweigh and redefine the estimand of interest based on functions of the conditional support for the respective interventions. With these functions, we obtain the desired dose-response curve in areas of enough support, and otherwise a meaningful estimand that does not require the positivity assumption. We develop $g$-computation type plug-in estimators for this case. Those are contrasted with g-computation estimators which are applied to continuous interventions without specifically addressing positivity violations, which we propose to be presented with diagnostics. The ideas are illustrated with longitudinal data from HIV positive children treated with an efavirenz-based regimen as part of the CHAPAS-3 trial, which enrolled children $<13$ years in Zambia/Uganda. Simulations show in which situations a standard g-computation approach is appropriate, and in which it leads to bias and how the proposed weighted estimation approach then recovers the alternative estimand of interest.
\end{abstract}

\begin{mdframed}[backgroundcolor=red!10, linecolor=black!50]
{\small
The published version of this working paper can be cited as follows:\\[0.25cm]
Schomaker, M., McIlleron, H., Denti, P., Díaz, I.\\
\textit{Causal Inference for Continuous Multiple Time Point Interventions}\\
Statistics in Medicine (2024); in press
}
\end{mdframed}

\section{Introduction}\label{sec:introduction}

Causal inference for multiple time-point interventions has received considerable attention in the literature over the past few years: if the intervention of interest is binary, popular estimation approaches include inverse probability of treatment weighting approaches (IPTW) \cite{Robins:2000}, g-computation estimators \cite{Robins:1986, Bang:2005} and longitudinal targeted maximum likelihood estimators (LTMLE) \cite{vanderLaan:2011}, among others. For IPTW, the treatment and censoring mechanisms need to be estimated at each time point, parametric g-computation requires fitting of both the outcome and confounder mechanisms, sequential g-computation is based on the iterated outcome regressions only, and LTMLE needs models for the outcome, censoring, and treatment mechanisms, iteratively at each point in time. All of those methods are relatively well-understood and have been successfully applied in different fields (e.g., in \cite{Schnitzer:2014b, Schomaker:2019, Baumann:2021, Bell-Gorrod:2020, Cain:2016, Kreif:2017}).

However, suggestions on how to estimate treatment effects of variables that are continuous and measured at multiple time points are limited. This may, however, be of interest to construct causal dose-response curves (CDRC). For example, in pharmacoepidemiology, one may be interested in how counterfactual outcomes vary for different dosing strategies for a particular drug, see Section \ref{sec:motivation}.

Early work on continuous interventions includes the seminal paper of Robins, Hernán and Brumback \cite{Robins:2000} on inverse probability weighting of marginal structural models (MSMs). For a single time point, this requires the estimation of stabilized weights based on the conditional density of the treatment, given the confounders. This density may be estimated with parametric regression models, such as linear regression models. The MSM, describing the dose-response relationship (i.e., the CDRC), can then be obtained using weighted regression. The suggested approach can also be used for the longitudinal case, where stabilized weights can be constructed easily, and one may work with a working model where, for example, the effect of cumulative dose over all time points on the response is estimated.

There are several other suggestions for the point treatment case (i.e., a single time point): for example, the use of the generalized propensity score (GPS) is often advocated in the literature \cite{Hirano:2004}. Similar to the MSM approach described above, both the conditional density of the treatment, given the confounders, and the dose-response relationship have to be specified and estimated. Instead of using stabilized weights, the GPS is included in the dose-response model as a covariate. To reduce the risk of bias due to model mis-specification, it has been suggested to incorporate machine learning (ML) in the estimation process \cite{Kreif:2015}. However, when combining GPS and ML approaches, there is no guarantee for valid inference, i.e., related confidence intervals may not achieve nominal coverage \cite{vanderLaan:2011}. Doubly robust (DR) approaches allow the integration of ML while retaining valid inference. As such, the DR-estimator proposed by Kennedy et al. \cite{Kennedy:2017} is a viable alternative to both MSM and GPS approaches. It does not rely on the correct specification of parametric models and can incorporate general machine learning approaches. As with other, similar, approaches \cite{Westling:2020} both the treatment and outcome mechanisms need to be modelled. Then a pseudo-outcome based on those two estimated nuisance functions is constructed, and put into relationship with the continuous intervention using kernel smoothing. Further angles in the point treatment case are given in the literature \cite{Zhang:2016, Galvao:2015, vanderWeele:2011, Diaz:2013}.

There are fewer suggestions as to how to estimate CDRCs for multiple time point interventions. As indicated above, one could work with inverse probability weighting of marginal structural models and specify a parametric dose-response curve. Alternatively, one may favour the definition of the causal parameter as a projection of the true CDRC onto a specified working model \cite{Neugebauer:2007}. Both approaches have the disadvantage that, with mis-specification of the dose-response relationship or an inappropriate working model, the postulated curve may be far away from the true CDRC. Moreover, the threat of practical positivity violations is even more severe in the longitudinal setup, and working with inverse densities – which can be volatile – remains a serious concern \cite{Goetgeluk:2008}. It has thus been suggested in the literature to avoid these issues by changing the scientific question of interest, if meaningful, and to work with alternative definitions of causal effects.

For example, Young et al. \cite{Young:2014} consider so-called modified treatment policies, where treatment effects are allowed to be stochastic and depend on the natural value of treatment, defined as the treatment value that would have been observed at time $t$, had the intervention been discontinued right before $t$. A similar approach relates to using representative interventions, which are stochastic interventions that maintain a continuous treatment within a pre-specified range \cite{Young:2019}. Diaz et al. \cite{Diaz:2020} present four estimators for longitudinal modified treatment policies (LMTPs), based on IPTW, g-computation and doubly robust considerations. These estimators are implemented in an R-package (\texttt{lmtp}). An advantage of those approaches is that the positivity assumption can sometimes be relaxed, depending on how interventions are being designed. Moreover, the proposed framework is very general, applicable to longitudinal and survival settings, and one is not forced to make arbitrary parametric assumptions if the doubly robust estimators are employed. It also avoids estimation of conditional densities as it recasts the problem at hand as a classification procedure for the density ratio of the post-intervention and natural treatment densities. A disadvantage of the approach is that it does not aim at estimating the CDRC, which may, however, relate to the research question of interest, see Section \ref{sec:motivation} below.

In this paper, we are interested in counterfactual outcomes after intervening on a continuous exposure (such as drug concentration) at multiple time points. For example, we may be interested in the probability of viral suppression after one year of follow-up, had HIV-positive children had a fixed concentration level of efavirenz during this year. In this example, it would be desirable to estimate the true CDRC as closely as possible, to understand and visualize the underlying biological mechanism and decide what preferred target concentrations should be. This comes with several challenges: most importantly, violations of the positivity assumption are to be expected with continuous multiple time-point interventions. If this is the case, estimands that are related to modified treatment policies or stochastic interventions, as discussed above, can be tailored to tackle the positivity violation problem, but redefine the question of interest. In pharmacoepidemiology and other fields this may be not ideal, as interpretations of the true CDRC are of considerable clinical interest, for example to determine appropriate drug target concentrations.

We propose g-computation based approaches to estimate and visualize causal dose-response curves. This is an obvious suggestion in our setting because developing a standard doubly robust estimator, e.g. a targeted maximum likelihood estimator, is \textit{not} possible as the CDRC is not a pathwise-differentiable parameter; and developing non-standard doubly robust estimators is not straightforward in a multiple time-point setting.
Our suggested approach has two angles:
\begin{enumerate}[i)]
\item As a first step, we simply consider computing counterfactual outcomes for multiple values of the continuous intervention (at each time point) using standard parametric and sequential g-computation. We evaluate with simulation studies where this ``naive'' estimation strategy can be successful and useful, and where not; and which diagnostics may be helpful in judging its reliability. To our knowledge, this standard approach has not been evaluated in the literature yet.
\item We then define regions of low support through low values of the conditional treatment density, evaluated at the intervention trajectories of interest. Our proposal is to redefine the estimand of interest (i.e., the CDRC) only in those regions, based on suitable weight functions. Such an approach entails a compromise between identifiability and interpretability: it is a tradeoff between estimating the CDRC as closely as possible [as in i)], at the risk of bias due to positivity violations and minimizing the risk of bias due to positivity violations, at the cost of redefining the estimand in regions of low support. We develop a g-computation based plug-in estimator for this weighted approach.
\end{enumerate}

We introduce the motivating question in Section \ref{sec:motivation}, followed by the theoretical framework in Section \ref{sec:framework}. After presenting our extensive Monte-Carlo simulations (Section \ref{sec:simulation}), we analyze the illustrative data example in Section \ref{sec:data_analysis}. We conclude in Section \ref{sec:discussion}.

\section{Motivating Example}\label{sec:motivation}

Our data comes from \textit{CHAPAS-3}, an open-label, parallel-group, randomised trial (CHAPAS-3) \cite{Mulenga:2016}. Children with HIV (aged 1 month to 13 years), from four different treatment centres (one in Zambia, three in Uganda) were randomized to receive one out of 3 different antiretroviral therapy (ART) regimens, given as fixed-dose-combination tablets. Each regimen consisted of 3 drugs. Every child received lamivudine (first drug), and either nevirapine or efavirenz (second drug), which was chosen at the discretion of the treating physician (and based on age). The third drug was randomly assigned (1:1:1) and either stavudine, zidovudine, or abacavir. The primary endpoint in the trial were adverse events, both clinical (grade 2/3/4) and laboratory (confirmed grade 3, or any grade 4).

Several substudies of the trial explored pharmacokinetic aspects of the nonnucleoside reverse transcriptase inhibitors (NNRTI) efavirenz and nevirapine (the second, non-randomized drug), in particular the relationship between the respective drug concentrations and elevated viral load, i.e. viral failure \cite{Bienczak:2016, Bienczak:2017}. Our analyses are motivated by these substudies and are based on the subsets of children, who received efavirenz (that is, 125 out of 478 patients). We use the same data as Bienczak et al. \cite{Bienczak:2016}.

In the trial, efavirenz dose was recommended to be based on weight using 200 mg for those weighing 10-13.9 kg, 300 mg for 14-19.9 kg, 400 mg for 20-34.9 kg, and 600 mg (the adult dose) above 34.9 kg. While children may receive the same dose, EFV concentrations vary individually and depend on the child's metabolism (that is, specifically the single nucleotide polymorphisms in the CYP2B6 gene encoding the key metabolizing enzyme).
Too low concentrations reduce antiviral activity of the drug and thus lead to viral failure (a negative outcome, which typically leads to a change in drug regimen). This is the reason why one is interested in the minimum concentration that is still effective against viral replication; or, more generally in the range of concentrations that should be targeted.

Our analysis evaluates the relationship between EFV concentrations (not doses) and viral failure, over a follow-up period of 84 weeks. We are specifically interested in the counterfactual probability of viral load (VL) $>100$ copies/ml at 84 weeks if children had concentrations (12/24h after dose) of $x$ mg/L at each follow-up visit, where $x$ ranges from $0$ to $6$ mg/L. This question translates into the longitudinal causal dose-response curve (CDRC), which --in our example--, is actually a concentration-response curve. That is, we want to know how the probability of failure varies for different concentration trajectories.

However, drawing a particular CDRC is challenging for the following reasons: (i) the form of the curve should be flexible and as close as possible to the truth; (ii) there exist time-dependent confounders (e.g., adherence, weight) that are themselves affected by prior concentration levels, making regression an invalid method for causal effect estimation \cite{Hernan:2020}; (iii) with long follow-up and moderate sample size, and given the continuous nature of the concentration variables, positivity violations are an issue of concern that have to be addressed.

More details on the data analysis are given in Section \ref{sec:data_analysis}.

\section{Framework}\label{sec:framework}

\subsection{Notation}\label{sec:framework_notation}
We consider a longitudinal data setup where at each time point $t$, $t=0,1,\ldots,T$, we measure the outcome $Y_t$, a continuous intervention $A_t$ and covariates $L^j_t$, $j=1,\ldots,q$ for $i = 1, \ldots, n$ individuals. We denote  $\mathbf{L_0}=\{L^1_0,\ldots,L^{q_0}_0\}$ as ``baseline variables'' and $\mathbf{L}_t=\{L^1_t,\ldots,L^q_t\}$ as follow-up variables, with $q,q_0 \in \mathbb N$. The intervention and covariate histories of a unit $i$ (up to and including time $t$) are $\bar{A}_{t,i}=(A_{0,i},\ldots,A_{t,i})$ and $\bar{L}^s_{t,i}=(L^s_{0,i},\ldots,L^s_{t,i})$, $s=1,...,q$, $i=1,\ldots,n$, respectively. The observed data structure is
\begin{eqnarray*}
O = (\mathbf{L}_0,A_0,Y_0,\mathbf{L}_1,A_1,Y_1,\ldots,\mathbf{L}_{T},A_{T},Y_{T}) \,.
\end{eqnarray*}

That is, we consider $n$ units, followed-up over $T$ time points where at each time point we work with the order $L_t \rightarrow A_t \rightarrow Y_t$.

We are interested in the counterfactual outcome $Y_{t,i}^{\bar{a}_{t}}$ that would have been observed at time $t$ if unit $i$ had received, possibly contrary to the fact, the intervention history $\bar{A}_{t,i}=\bar{a}_t$, $a_{t,i} \in \mathbb{R}$. For a given intervention $\bar{A}_{t,i}=\bar{a}_t$, the counterfactual covariates are denoted as $\bar{\mathbf{L}}_{t,i}^{\bar{a}_{t}}$. We use $\mathbf H_t$ to denote the history of all data up to before $A_t$.

\subsection{Estimands}\label{sec:framework_estimands}

\subsubsection{Estimands for one time point}
\begin{figure}[ht!]
\begin{center}
\subfloat[CDRC for one time point]{\label{figure:CDRC_single}\includegraphics[scale=0.3]{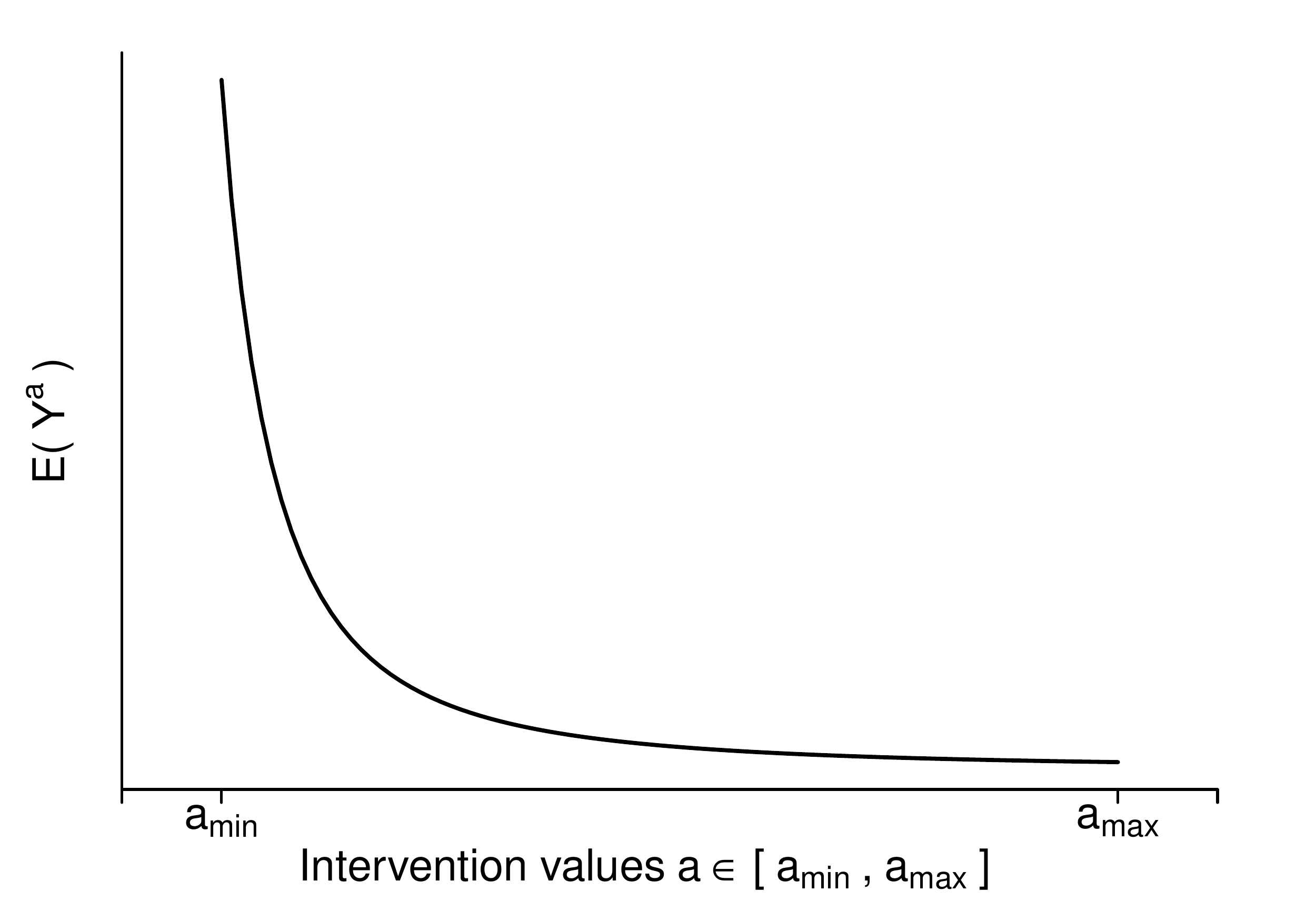}\hfill}\hfill
\subfloat[Possible CDRC visualization \\-- if intervention values are constant over time]{\label{figure:CDRC_multiple_1}\includegraphics[scale=0.3]{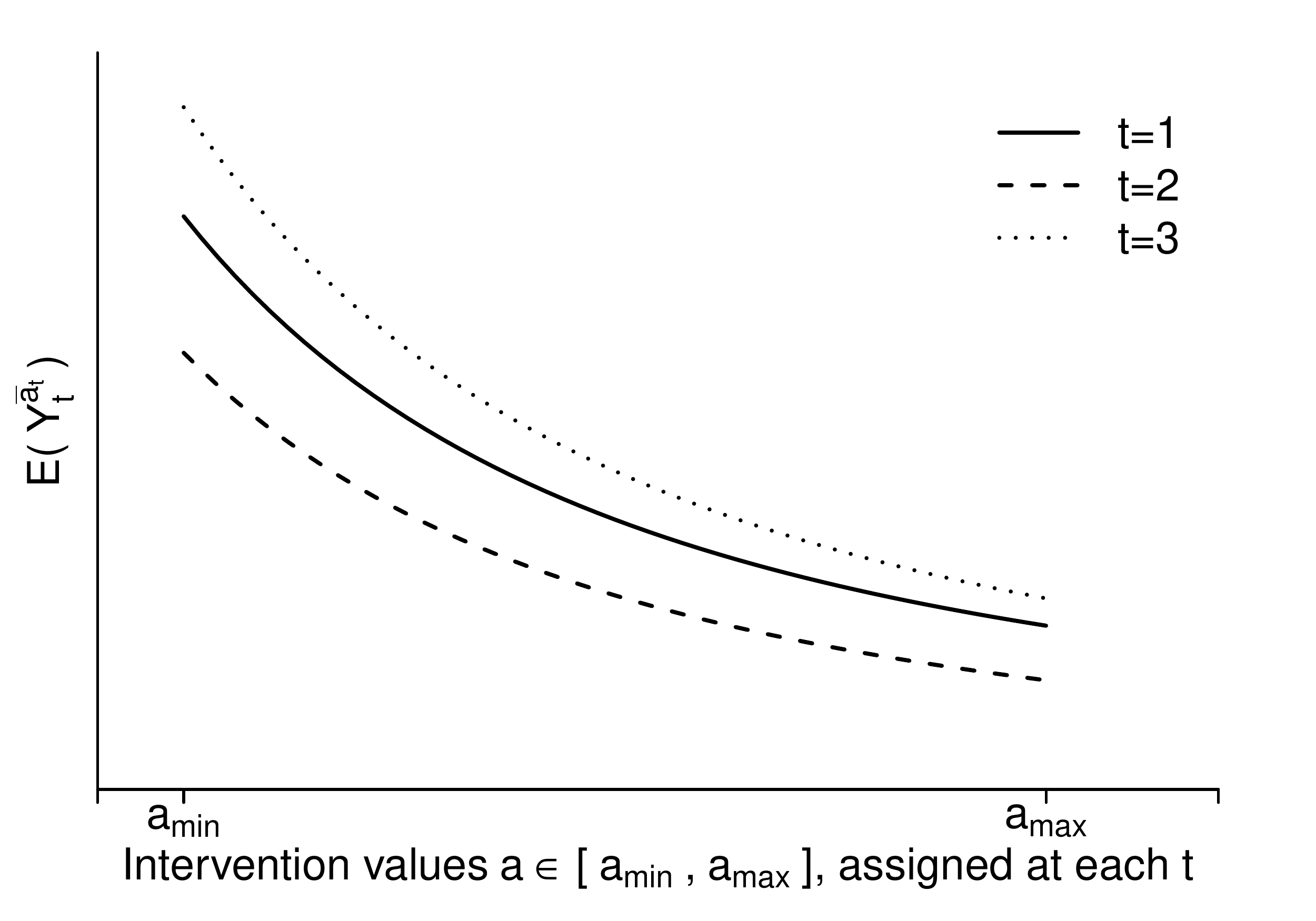}\hfill}\\
\subfloat[Possible CDRC visualization \\-- if intervention values vary over time (or for survival data)]{\label{figure:CDRC_multiple_2}\includegraphics[scale=0.3]{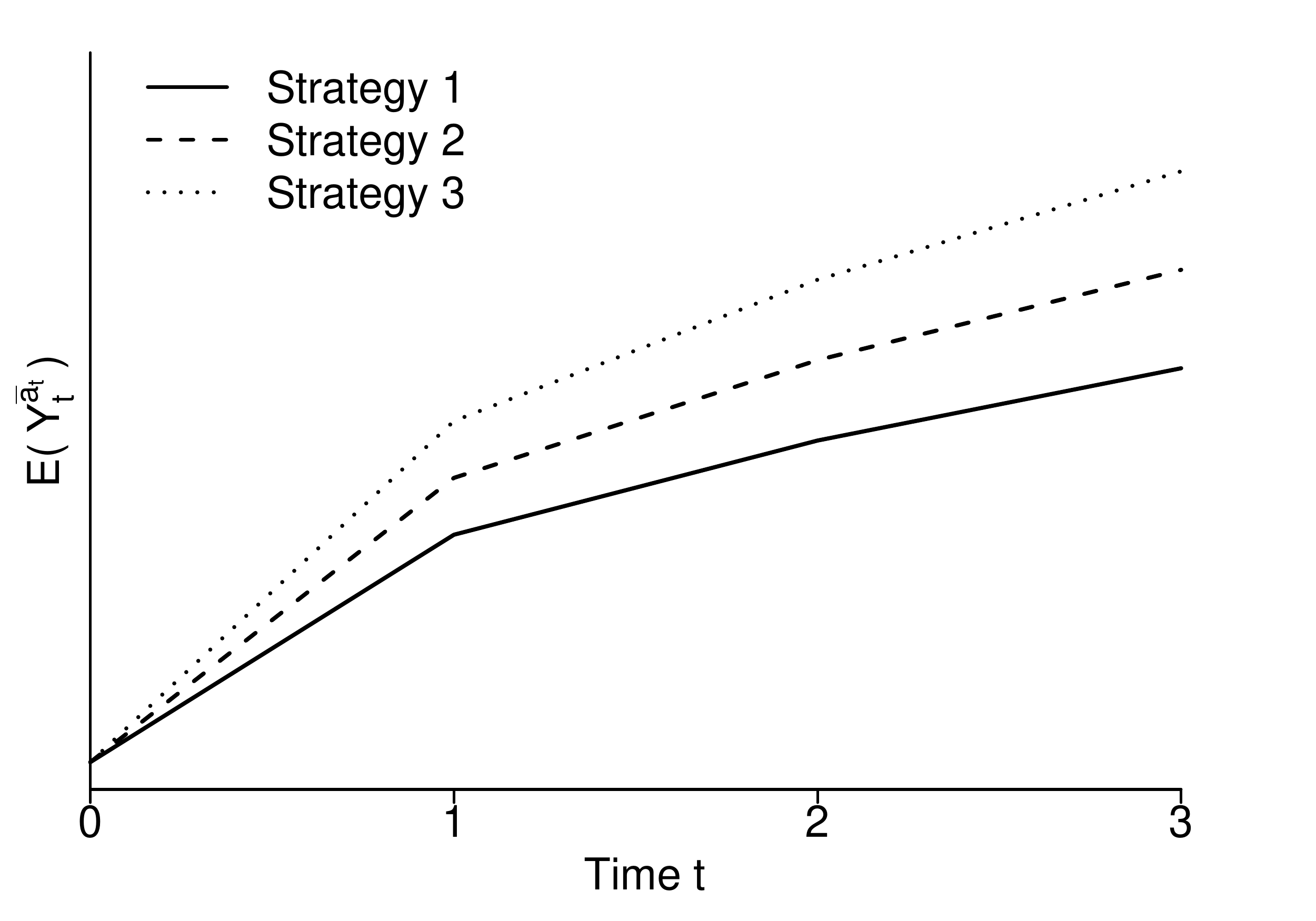}}
\caption{Considerations for causal dose-response curves (CDRCs)}\label{figure:CDRC_considerations}
\end{center}
\end{figure}

In order to illustrate the ideas, we first consider a simple example with only one time point where $O=(\mathbf L,A,Y)$. Suppose we observe an independent and identically distributed sample $O=(O_1,\ldots, O_n)$, where $O$ has support $supp({O}) = \{\mathcal{L} \times \mathcal{A} \times \mathcal{Y}\}$. Our estimand of interest is the causal dose-response curve:
\begin{flalign}\label{eqn:estimand1_time1}
\text{\bf Estimand 1:} \quad & m: a \mapsto E(Y^a) \quad \forall a \in \mathcal{A} \,, &
\end{flalign}
where $\mathcal{A} = {supp}(A) := \{ a \in A \,:\, f(a) \neq 0\}$. Assume $\mathbf{L}$ has density $p_0(\mathbf{l})$ with respect to some dominating measure $\nu(\mathbf{l})$. The dose-response curve  (\ref{eqn:estimand1_time1}) can be identified \cite{Kennedy:2017} as
\begin{align}
    m(a) &=  E_{\mathbf{L}}[E(Y^a\mid A=a, \mathbf{L})]
    = E_{\mathbf{L}}[E(Y\mid A=a, \mathbf{L})] \nonumber\\
    &= \int_{\mathbf{l} \in \mathcal{L}} E(Y\mid A=a, \mathbf{L}=\mathbf{l})p_0(\mathbf{l}) d\nu(\mathbf{l})\,, \label{eqn:gcomp}
\end{align}
where the first equality follows by the law of iterated expectation, the assumption that $Y^a$ is independent of $A$ conditional on $\mathbf{L}$ (conditional exchangeability) and positivity (see below). The second equality follows because $Y^a=Y$ in the event of $A=a$ (consistency). An example of such a curve for a particular interval $[a_{\min}, a_{\max}]$ is given in Figure \ref{figure:CDRC_single}.

This quantity is undefined if there is an intervention level $a \in \mathcal{A}$ such that the conditional density function $g(a\mid \mathbf{l})$ is zero for some $\mathbf{l}$ with $p_0(\mathbf{l})>0$. The \textit{strong positivity assumption} reflects this consideration by requiring that $\inf_{{a} \in {\mathcal{A}}} g({a}\mid \mathbf{l}) > 0$ almost everywhere \cite{Petersen:2012}. This assumption may be weakened by making additional parametric model assumptions or restricting the range of $a$.
In this paper, we however address violations of the positivity assumption through a redefinition of the estimand of interest, where we replace the marginal distribution function of $\mathbf{L}$ in (\ref{eqn:gcomp}) by a user-given distribution function. That is, we instead target 
\begin{flalign}\label{eqn:estimand2_time1_general}
     m_w(a) = \int E(Y\mid A=a, \mathbf{L}=\mathbf{l})w(a,\mathbf{l}) p_0(\mathbf{l}) d\nu(\mathbf{l})
\end{flalign}
for some weight function $w(a,\mathbf{l})$. This is essentially a weighted average of the conditional dose-response curve $E(E(Y^a \mid \mathbf{L})\,w(a,\mathbf{L}))$.

To illustrate the construction of a meaningful weight function for (\ref{eqn:estimand2_time1_general}), consider two extreme cases: $w(a,\mathbf{l}) = 1$, and $w(a,\mathbf{l}) = g(a\mid \mathbf{l})/g(a)$, where $g(a)$ is the marginal density of $A$. Under the first case, we have $m_w(a)=m(a)$, which is equal to the dose-response curve (estimand 1) whenever the positivity assumption holds. Under $w(a,\mathbf{l}) = g(a\mid \mathbf{l})/g(a)$, we have $m_w(a)=E(Y\mid A=a)$, which is not a causal quantity (unless exchangeability is assumed) but it does not require the positivity assumption. We therefore propose to use a function $w(a,\mathbf{l})$ such that $w(a,\mathbf{l}) = 1$ in areas of $\{\mathcal{A}, \mathcal{L}\}$ that have good support, and $w(a,\mathbf{l}) = g(a\mid \mathbf{l})/g(a)$ in areas of $\{\mathcal{A}, \mathcal{L}\}$ that have low support. By ``good support'', we mean that the positivity assumption is met in the sense of $g(a\mid \mathbf{l}) > c > 0$, i.e. bounded away from zero.  Thus, one possible function to use is
\begin{equation}\label{eqn:w_1}
w(a,\mathbf{l}) =
\begin{cases}
1 & \text{ if } {g(a\mid \mathbf{l})} > c \\
\frac{g(a\mid l)}{g(a)} & \text{ otherwise. }\\
\end{cases}
\end{equation}

Alternatively, one may use $g(a\mid \mathbf{l})/g(a) > c > 0$ to define good support. Using (\ref{eqn:w_1}) in (\ref{eqn:estimand2_time1_general}) targets the desired dose-response curve under enough support and avoids reliance on positivity otherwise. Formally,

\noindent\text{\bf Estimand 2 is then given by:}
\begin{eqnarray}
\label{eqn:estimand2_time1_specific}
m_{w}(a) & = &
E(Y^a \mid g(a\mid L) > c) P(g(a\mid L) > c) \nonumber\\ && + E(Y^a w(a\mid L) \mid g(a\mid L) \leq c) P(g(a\mid L) \leq c) \quad\quad \forall a \in \mathcal{A}
\end{eqnarray}
For fixed $a$, this is taking a weighted average of the counterfactuals $Y^a$, where the weights are equal to one for units in the population for whom positivity $g(a\mid L) > c$ holds, and the weights are $w(a\mid L) < c /g(a)$ for those for whom it does not. Whenever $c/g(a)<1$, this simply downweights observations that rely more on extrapolation so that they are not as influential in the estimators, and imply an interpretation as outlined below otherwise. \\

\noindent{\textbf{Interpretation.}} To understand the implication and interpretation of using this estimand, consider our motivating data example: first, note that both $m(a)$ and $m_{w}(a)$ are undefined outside the support region of $A$. For instance, in our study the CDRC is undefined for negative concentration values and biologically implausible concentration values of $>40$ mg/L. Suppose we are interested in $E(Y^a)$ for $a=0.5$ mg/L: for all patients with covariate regions $\mathbf{l}$ that have $g(0.5 \mid \mathbf{l}) > c$ we stick to the CDRC m(a), such that we obtain $E(Y^a)$; but for those where this does not hold, maybe because they are ultraslow metabolizers who will not clear the drug fast enough to ever achieve $0.5$ mg/L under full adherence (i.e.. $g(0.5 \mid ultraslow, adherent) = 0$), we target $E(Y|a)$. This means that we do not require positivity or rely on parametric extrapolations in poor support regions (where the intervention seems ``unrealistic''); we rather use the present associations to allow individual concentration trajectories that lead to $E(Y|0.5)$ for those patient groups. Thus, the proposed estimand (\ref{eqn:estimand2_time1_specific}) offers a tradeoff between identifiability and interpretability: that is, a tradeoff between estimating the CDRC as closely as possible, at the risk of bias due to positivity violations because of the continuous intervention; and minimizing the risk of bias due to positivity violations, at the cost of redefining the estimand. More details on possible interpretations are given in Sections \ref{sec:framework_multiple_time_points} and \ref{sec:framework_both_estimands}.\\

\noindent{\textbf{Choice of c.}}
As in the case for binary interventions, when truncating the propensity score for inverse probability weighting or targeted maximum likelihood estimation, one may use rules of thumbs and simulation evidence to decide for $c$. This is because we want the conditional treatment density to be bounded away from zero, to avoid negative effects of (near-)positivity violations. Possible ad-hoc choices are $0.01$ and $5 / (\sqrt(n) \ln n/5 )$ \cite{Gruber:2022}, though we argue below that multiple $c's$ may be selected and presented based on diagnostics, see Section \ref{sec:framework_positivity}.   \\

\noindent\textbf{Estimands under different weight functions.} In principal, one could construct other weight functions too. An obvious choice would be to simply use the marginal treatment density $g(a)$ as a weight. Such a weight choice has been motivated in the context of estimands that are defined through parameters in a marginal structural working model for (continuous summaries of) longitudinal binary interventions \cite{Petersen:2014b}. In this case, greater weight is given to interventions with greater marginal support; and the more support there is for each possible intervention choice, the estimand will be closer to the CDRC. While intuitively it may make sense to rely more on interventions that are more often observed in the data, this approach does not directly address positivity violations; in the example above, there may be enough patients with concentrations close to $0.5$ mg/L overall, but not among ultraslow metabolizers, which is the issue to be addressed.

\subsubsection{Estimands for Multiple Time Points}\label{sec:framework_multiple_time_points}
To illustrate our proposed concepts for multiple time points, consider data for two time points first: ($\mathbf{L}_0, A_0, Y_0, \mathbf{L}_1, A_1, Y_1$). We are interested in $E(Y_1^{A_0=a_0, A_1=a_1})$. In principle, we may be interested in any $(a_0, a_1)$ within the support region $\{\mathcal{A}_1 \times \mathcal{A}_2\}$. Practically, it may be possible that we only care about interventions $(a_0, a_1)$ for which $a_0=a_1 > 0$ (as in the motivating example), or $a_0 < a_1$: in this case, we may restrict the estimand to the respective region $\{\mathcal{A}_1^{\ast} \times \mathcal{A}_2^{\ast}\} \subset \{\mathcal{A}_1 \times \mathcal{A}_2\}$. Similar to a single time point there may be the situation where $g(a_1) > 0$, $g(a_1, a_0)>0$ but $g(a_1 \mid \mathbf{l}_1, a_0, y_0, \mathbf{l}_0) \approx 0$. This corresponds to a situation where there is little support for the intervention value $a_1$ given that we already intervened with $a_0$ and given the covariate history; for regions, where this is the case, we address the respective positivity violations by redefining the estimand.\\

\noindent\textbf{A) Causal Dose-Response-Curve.} If there are multiple time points, the CDRC is
\begin{flalign}\label{eqn:estimand_general}
\text{\bf Estimand 1:} \quad & m_t: \bar{a}_t \mapsto E(Y^{\bar{a}_t}_t) \quad \quad \forall \bar{a}_t \in \mathcal{\bar{A}}_t, \forall t \in \mathcal{T} \,,&
\end{flalign}

where $\mathcal{T} = \{0,1,\ldots,T\}$ and $\bar{\mathcal{A}}_t = \{\mathcal{A}_1 \times \ldots \times \mathcal{A}_t\}$. The estimand (\ref{eqn:estimand_general}) can, in principle, be identified through various ways. A possible option under the assumptions of sequential conditional exchangeability, consistency and positivity \cite{vanderLaan:2012} --as defined below-- is the sequential g-formula (also known as the iterated conditional expectation representation \cite{Bang:2005}):
\begin{eqnarray}\label{eqn:seq_g_formula}
\mathbb{E}(Y_t^{\bar{a}_t}) &=& \mathbb{E}(\,\ldots\mathbb{E}(\,\mathbb{E}(Y_t|{{A}}_{t}={a}_{t}, \mathbf{{H}}_t) | {{A}}_{t-1}={a}_{t-1}, \mathbf{{H}}_{t-1}\,)\ldots|{A}_{0}={a}_{0}, \mathbf{{H}}_{0}\,)\,)\,,
\end{eqnarray}

In the above expression, $\bar{Y}_{t-1}$ is part of $\mathbf{\bar{L}}_t$, and thus $\mathbf{H}_t$. Because $A_t$ is continuous, a \textit{strong positivity} requirement corresponds to
\begin{equation}\label{eqn:positivity_longitudinal}
\inf_{a_t \in {\mathcal{A}}_t} g({a}_t\mid \mathbf{h}_t) > 0 \quad \text{whenever} \quad p_0(\mathbf l_{t}\mid A_{t-1} = a_{t-1}, \mathbf H_{t-1}=\mathbf h_{t-1})>0 \quad \forall t \in \mathcal{T}\,.
\end{equation}

This means, we require within the support region $\bar{\mathcal{A}}_t$ a positive conditional treatment density for each $a_t$, given its past. This strong assumption may be relaxed either under additional parametric modeling assumptions or by a restriction to some $\mathcal{\bar A}_t^{\ast} \subset \mathcal{\bar A}_t$ or by a redefinition of the estimand. We continue with the latter strategy.
Consistency in the multiple time-point case is the requirement that $Y^{\bar{a}_t}_t = Y_t$ if $\bar{A}_{t} = \bar{a}_{t}$ and $\bar{\mathbf{L}}_t^{\bar{a}_{t-1}}=\bar{\mathbf{L}}_{t}$ if $\bar{A}_{t-1} = \bar{a}_{t-1}$. With sequential conditional exchangeability we require the counterfactual outcome under the assigned treatment trajectory to be independent of the actually assigned treatment at time $t$, given the past: $Y^{\bar{a}_t}_t\coprod A_{t}|{H}_{t}$ for $t = 0,\ldots,T$.

\textit{Note:} Sometimes, we may want to visualize the CDRC graphically, e.g. by plotting $m_t(\bar a_t)$ for each $t$ (and stratified by $\mathbf{L_0}$, or a subset thereof, if meaningful) if the set of strategies is restricted to those that always assign the same value at each time point, see Figure \ref{figure:CDRC_multiple_1} for an illustration. Alternatively, we may opt to plot the CDRC as a function of $t$, for some selected strategies $\bar{a}_t$, see Figure \ref{figure:CDRC_multiple_2}. This may be useful if intervention values change over time, or for survival settings.\\

\noindent\textbf{B) Weighted Estimand.}
To link the above longitudinal g-computation formula (\ref{eqn:seq_g_formula}) to our proposals, it may be written in terms of the following recursion. Let $\tilde{m}_{t+1} = Y_t$. For $s=t,\ldots,0$ recursively define
\begin{equation}\label{eqn:CDRC_long_recursive}
\tilde{m}_s: (a_s, \mathbf h_s) \mapsto \int \tilde{m}_{s+1}(a_{s+1}, \mathbf h_{s+1}) \, d P_0(\mathbf l_{s+1}\mid A_s = a_s, \mathbf H_s=\mathbf h_s)\,,
\end{equation}
where $\int \tilde{m}_{s+1}(a_{s+1}, \mathbf h_{s+1}) \, d P_0(\mathbf l_{s+1}\mid A_s = a_s, \mathbf H_s=\mathbf h_s)$ = $E(\tilde{m}_{s+1}(a_{s+1}, \mathbf{H}_{s+1})|A_s=a_s,\mathbf{H}_s=\mathbf{h}_s)$.\\

Then, the counterfactual mean outcome $E(Y_t^{\bar{a}_t})$ is identified as $\mathbb{E}(\tilde{m}_0(a_0,h_0))$, which follows from the re-expression of (\ref{eqn:CDRC_long_recursive}) in terms of (\ref{eqn:seq_g_formula}) by recursively evaluating the integral; see Appendix \ref{sec:appendix_theory_CDRC_details} for details. The above recursive integral is well defined only if the positivity assumption (\ref{eqn:positivity_longitudinal}) is met. As for the single time-point case, we propose to address violations of the positivity assumption by targeting a modified identifying expression, which we express through the following recursive integral:
\begin{equation}\label{eqn:estimand_general_long}
\tilde{m}_{w,s}:(a_s, \mathbf h_s)\mapsto \int \tilde{m}_{s+1}(a_{s+1}, \mathbf h_{s+1}) w_s(a_{s+1}, \mathbf h_{s+1})d P_0(\mathbf l_{s+1}\mid A_s = a_s, \mathbf H_s=\mathbf h_s)\,.
\end{equation}
As before, if the weight function $w_s(a_{s+1}, \mathbf h_{s+1})$ is equal to one, the above expression can be used to define the actual CDRC (\ref{eqn:estimand_general}) through recursive evaluation of (\ref{eqn:estimand_general_long}). If, however, the weight function is equal to
\begin{equation}\label{eqn:w1_long}
    w_{s}(a_{s+1}, \mathbf h_{s+1}) = \frac{g_s(a_{s+1}\mid \mathbf h_{s+1})}{g_s(a_{s+1}\mid a_s, \mathbf h_s)},
\end{equation}
then the expression becomes
\begin{equation}\label{eqn:estimand_weighted_long}
\tilde{m}_{w,s}: (a_s, \mathbf h_s) \mapsto \int \tilde{m}_{s+1}(a_{s+1},\, \mathbf h_{s+1})d P_0(\mathbf l_{s+1}\mid A_{s+1} = a_{s+1}, A_s = a_s, \mathbf H_s=\mathbf h_s)
\end{equation}
by application of Bayes' rule. Intuitively, the above quantity does not remove confounding of the relation between $A_{s+1}$ and $Y_{s+1}$, because it conditions on $A_{s+1}=a_{s+1}$ rather than fixing (i.e., intervening on) $A_{s+1}=a_{s+1}$. However, the above expression does not require the positivity assumption to be well defined, as motivated further below. Generally speaking, applying the above weights iteratively over all time points (and all units) leads to
$$E(Y_t\mid \bar{a}_t),$$
see Appendix \ref{sec:appendix_theory_weights} for details.

For the longitudinal case, we follow a strategy similar to that considered for the single time-point case, and consider a compromise between satisfying the positivity assumption and adjusting for confounding. This compromise can be achieved by using weight functions that are $1$ in areas of good support, and are equal to (\ref{eqn:w1_long}) in areas of poor support:

\begin{equation}\label{eqn:w_1t}
w_{s}(a_{s+1},\mathbf h_{s+1},c) =
\begin{cases}
1 & \text{ if } g_s(a_{s+1}\mid \mathbf h_{s+1}) > c \,,\\
\frac{g_s(a_{s+1}\mid \mathbf h_{s+1})}{g_s(a_{s+1}\mid a_s, \mathbf h_s)} & \text{ if } g_s(a_{s+1}\mid \mathbf h_{s+1}) \leq c \text{ (and } g_s(a_{s+1}\mid a_s, \mathbf h_s) > c \text{)}
\end{cases}
\end{equation}

\noindent Using the weights (\ref{eqn:w_1t}) in (\ref{eqn:estimand_general_long}) has similar implications as for a single time point; that is, we obtain a weighted average of counterfactuals depending on the units' support for a given strategy $\bar{a}_t$. Formally, \noindent\text{\bf Estimand 2} equates to:

\begin{flalign}\label{eqn:estimand2_time_multiple}
m_{w,t}(\bar{a}_t)
 = \, &
E(Y_t^{\bar{a}_t} \mid \{g(a_s \mid \mathbf{H}_s) > c\}_{s=0}^t) \,  P(\{g(a_s \mid \mathbf{H}_s) > c\}_{s=0}^t) \,+ \nonumber\\
&  E(Y_t^{\bar{a}_t} \tilde{w}_t \mid \vee_{s=0}^t \{ g(a_s \mid \mathbf{H}_s) \leq c \}
) \,  P(\vee_{s=0}^t \{ g(a_s \mid \mathbf{H}_s) \leq c \}) \,  \quad \quad \quad\forall \bar{a}_t \in \mathcal{\bar{A}}_t, \forall t \in \mathcal{T}
\end{flalign}

where $\{g(a_s \mid \mathbf{H}_s) > c\}_{s=0}^t  = (g(a_t \mid \mathbf{H}_t) > c, \ldots, g(a_0 \mid \mathbf{H}_0) > c))$, $\vee_{s=0}^t \{ g(a_s \mid \mathbf{H}_s) \leq c \}$ indicates that at least one $g(a_s \mid \mathbf{H}_s) \leq c$, $w_t = w_{t}(a_{t+1},\mathbf h_{t+1},c)$, and the notation $\tilde{w}_t$ indicates that the weights have been applied up to and including time $t$. Practically, this means that

\begin{enumerate}[i)]
\item for units where there is enough conditional support for a strategy $\bar{a}_s$ in terms of $g_s(a_s \mid \mathbf{h}_s) > c $ for all time points $s$ up to and including time $t$, the weights are 1. For those, we still target the CDRC, i.e. $E(Y_t^{\bar{a}_t})$.
\item for units that do \textit{not} have enough conditional support for a strategy at all time points, the weights in (\ref{eqn:w_1t}), second row, are used. With this, intuitively we target $E(Y_t\mid \bar{a}_t)$ as outlined above.
\item if there exist units for which there is support for a strategy at some points, but not at others, the weights will be 1 at some time points, but not for all. The implication is that we target the CDRC as much as this possible and deviate from it only at the time points where it is necessary.
\end{enumerate}

Note that for the second estimand \textit{no positivity assumption} as defined in (\ref{eqn:positivity_longitudinal}) is required for identification because whenever a (near-)positivity violation is present in terms of $g_s(a_s \mid \mathbf{h}_s) \leq c$ the weights (\ref{eqn:w_1t}) redefine the estimand in a way such that the assumption is not needed.\\

If the denominator in (\ref{eqn:w_1t}) is very small, i.e. $\leq c$, one could replace it with a very small ad-hoc value. Alternatively, one may want to compromise in the sense of evaluating the conditional density until previous time points:

\begin{equation}\label{eqn:w_1t_alt}
w_{s}(a_{s+1},\mathbf h_{s+1},c) =
\begin{cases}
1 & \text{ if } g_s(a_{s+1}\mid \mathbf h_{s+1}) > c \,,\\
\frac{g_s(a_{s+1}\mid \mathbf h_{s+1})}{g_s(a_{s+1}\mid a_s, \mathbf h_s)} & \text{ if } g_s(a_{s+1}\mid \mathbf h_{s+1}) \leq c \text{ and } g_s(a_{s+1}\mid a_s, \mathbf h_s) > c \,,\\[6pt]
\frac{g_s(a_{s+1}\mid \mathbf h_{s+1})}{g_s(a_{s+1}\mid a_{s-1}, \mathbf h_{s-1})} & \text{ if } g_s(a_{s+1}\mid \mathbf h_{s+1}) \leq c \text{ and } g_s(a_{s+1}\mid a_s, \mathbf h_s) \leq c\\ & \text{ and }g_s(a_{s+1}\mid a_{s-1}, \mathbf h_{s-1}) > c \,,\\
\vdots & \vdots \\[2pt]
\frac{g_s(a_{s+1}\mid \mathbf h_{s+1})}{g_s(a_{s+1})} & \text{ otherwise }.\\
\end{cases}
\end{equation}

\noindent\textbf{Interpretation of the weighted estimand.} The weighted estimand can be interpreted as follows: those units which have a covariate trajectory that makes the intervention value of interest at $t$ not unlikely to occur (under the desired intervention before $t$), receive the intervention at $t$; all other units get different interventions that produce, on average, outcomes as we would expect among those who actually follow the intervention trajectory of interest (if the weight denominator is well-defined). Informally speaking, we calculate the CDRC in regions of enough support and stick to the actual research question, but make use of the marginal associations otherwise, if possible.

As a practical illustration for this interpretation, consider our motivating study: if, given the covariate and intervention history, it seems possible for a child to have a concentration level of $a$ mg/l, then we do indeed calculate the counterfactual outcome under $a$. For some patients however, it may be unlikely (or even biologically impossible!) to actually observe this concentration level: for example, children who are adherent to their drug regimen and got an appropriate drug dose prescribed, but are slow metabolizers will likely never be able to have very low concentration values. In this case, we do not consider the intervention to be \textit{feasible} and rather let those patients have individual concentration levels which generate outcomes that are typical for children with $a$ mg/l. In short, we calculate the probability of failure at time $t$ if the concentration level is set at $a$ for all patients where this seems ``feasible'', and otherwise to individual concentration trajectories that produce ``typical'' outcomes with $a$ mg/l.

The more complex cases of the weighted estimand reflect the fact that certain strategies become unlikely  only at certain time points: for example, patients who are adherent to their treatment and receive always the same dose will likely not have heavy varying  concentrations in their body; thus, a strategy that looks at the effect of a concentration jump during follow-up, e.g. (3,3,3,0) will only be unlikely at the fourth visit, but not before - and hence a deviation from the CDRC will only be needed at the fourth time point.

The interpretation of this weighted dose-response curve seems somewhat unusual; note however that this approach has the advantages that i) compared to a naive approach it does not require the positivity assumption, and ii) compared to LMTP's it both sticks to the actual research question as close as ``possible'' and does not require a positvity assumption for the chosen policies of interest. We explain below, in Section \ref{sec:framework_both_estimands}, why we believe, however, that the weighted estimand (estimand 2) should typically be presented together with the CDRC, i.e. estimand 1. \\

\subsection{Estimation} \label{sec:framework_estimation}
To develop an estimation strategy, it is useful to see that we can reexpress our target quantity (\ref{eqn:estimand_general_long}) in terms of the iterated weighted conditional expectation representation
\begin{eqnarray}\label{eqn:seq_g_formula_weighted}
\tilde{m}_{w,t} &=& \mathbb{E}(\mathbb{E}(\ldots\mathbb{E}(\,\mathbb{E}(Y_t w_t|\bar{{A}}_{t}=\bar{a}_{t}, \mathbf{\bar{L}}_t) \, w_{t-1} | \bar{{A}}_{t-1}=\bar{a}_{t-1}, \mathbf{\bar{L}}_{t-1}\,)\ldots)\, w_0|{A}_{0}={a}_{0}, \mathbf{{L}}_{0}\,)\,)\,,
\end{eqnarray}
where $w_s = w_s(a_{s+1}, \mathbf h_{s+1})$; see Appendix \ref{sec:appendix_theory_estimand2_details} for details. This means that estimand 2 can be written as a series of iterated weighted outcomes. An alternative representation, based on a weighted parametric $g$-formula is given in Appendix \ref{sec:appendix_theory_parametric_gformula}.

For estimand 2, one can build a substitution estimator based on either (\ref{eqn:seq_g_formula_weighted}) or (\ref{eqn:gformula_parametric_weighted}).  That is, we use either a parametric or sequential g-formula type of approach where the (iterated) outcome is multiplied with the respective weight at each time point, regressed on its past, and recursively evaluated over time for a specific intervention $\bar{a}_t$. More specifically, a substitution estimator of (\ref{eqn:seq_g_formula_weighted}) can be constructed as follows: i) estimate the innermost expectation, i.e. multiply the estimated weights (\ref{eqn:w_1t}) with the outcome at time $t$; then, ii) intervene with the first intervention trajectory of interest $\bar{a}_t^{(1)}$;  iii) predict the weighted outcome under this intervention; iv) estimate the second innermost expectation by regressing the product of the prediction from iii) and the respective weight on its past; v) then, intervene with $\bar{a}_{t-1}^{(1)}$ and obtain the predicted weighted outcome under the intervention. These steps can be repeated until $t=0$, and for each intervention trajectory of interest.

As the weighted (iterated) outcomes may possibly have a skewed, complex distribution (depending on the weights), we advocate for the use of (\ref{eqn:seq_g_formula_weighted}) as a basis for estimation. This is because estimating the expectation only is typically easier than estimating the whole conditional distribution, as needed for (\ref{eqn:gformula_parametric_weighted}). In many applications, a data-adaptive estimation approach may be a good choice for modeling the expectation of the weighted outcomes.

For the estimates of the weighted curve to be consistent, both the expected weighted (iterated) outcomes and the weights need to be estimated consistently.

Note that for $t=0$, we suggest to calculate a weighted mean, instead of estimating the weighted expected outcome data-adaptively. This is because the former is typically more stable when $Y_0$ follows some standard distribution, but $w_0$ does not. It is a valid strategy as standardizing with respect to the pre-intervention variables $L_0$ and then calculating the weighted mean of the outcome is identical to standardizing the weighted outcome with respect to $L_0$, and then calculating the mean. Another option for $t=0$ is to facilitate step 6 before step 5 and then create a stacked dataset of the counterfactual outcomes for all intervention values $a_0$, and then fit a weighted regression of $Y_0$ on $A_0$, based on the stacked weight vectors. The fitted model can then be used to predict the expected outcome under all $a_0$. This strategy is however not explored further in this paper.

The algorithm in Table \ref{tab:estimation_gcomp_multiple} presents the detailed steps for a substitution estimator of (\ref{eqn:seq_g_formula_weighted}). The algorithm is implemented in the $R$-package mentioned below, and further illustrated in Appendix \ref{sec:appendix_code_algorithm}.

\begin{table}[ht!]
\caption{Algorithm for estimating estimand 2 at time $t$ }\label{tab:estimation_gcomp_multiple}
\small
\renewcommand{\arraystretch}{1.2}
\begin{tabular}{p{0.0925\textwidth}p{0.875\textwidth}}
\hline
Step 0a: & Define a set of interventions $\bar{\mathcal{A}}_t$, with $|\bar{\mathcal{A}}_t|=n_a$ and $\bar{A}^{(j)}_t$ is the $j^{\text{th}}$ element of $\bar{\mathcal{A}}_t$, $j=1,\ldots,n_a$. \\
Step 0b: & Set $\tilde{Y}_t = Y_t$.\\
\multicolumn{2}{l}{{\underline{\textbf{For $\mathbf{s = t,\ldots,1}$:}}}} \\
Step 1a: & Estimate the conditional density $g(A_s|\mathbf{H}_s)$, see also footnote 1. \\
Step 1b: & Estimate the conditional density $g(A_s|A_{s-1},\mathbf{H}_{s-1})$, see also footnote 1. \\
Step 2: &  Set $a_s=a_s^{(1)}$, where $a_s^{(1)}$ is the $s^{\text{th}}$ element of intervention $\bar{A}^{(1)}_t \in \bar{\mathcal{A}}_t$. \\
Step 3a: & Plug in $a_s^{(1)}$ into the estimated densities from step 1, to calculate $\hat{g}(a_s^{(1)}|\mathbf{h}_s)$ and $\hat{g}(a_s^{(1)}|a_{s-1}^{(1)},\mathbf{h}_{s-1})$.  \\
Step 3b: & Calculate the weights $w_s(a_s^{(1)},c)$ from (\ref{eqn:w_1t_alt}) based on the estimates from 3a.\\
&  If $\hat{g}(a_s^{(1)}|a_{s-1}^{(1)},\mathbf{h}_{s-1})\leq c$, estimate $g(a_s^{(1)}|a_{s^{\ast}}^{(1)},\mathbf{h}_{s^{\ast}})$ as required by the definition of (\ref{eqn:w_1t_alt}) for $s^{\ast}=s-2,\ldots,0$. \\
Step 4: & Estimate $\mathbb{E}(w_s(a_s^{(1)},c) \, \tilde{Y}_s|{A}_s, \mathbf{H}_s)$, see also footnote 2. \\
Step 5: & Predict $\tilde{Y}_{s-1} = \hat{\mathbb{E}}(\hat{w}_s(a_s^{(1)},c) \, \tilde{Y}_s|\bar{A}_s=\bar{a}_s^{(1)}, \mathbf{l}_s)$ based on the fitted model from step 4 and the given intervention $\bar{a}_s^{(1)}$. \\
\multicolumn{2}{l}{{\underline{\textbf{For $\mathbf{t = 0}$:}}}}  \\
Step 1a: & Estimate the conditional density $g(A_0|\mathbf{L}_0)$, see also footnote 1. \\
Step 1b: & Estimate the conditional density $g(A_0)$, see also footnote 1. \\
Step 2: &  Set $a_0=a_0^{(1)}$. \\
Step 3a: & Calculate $\hat{g}(a_0^{(1)}|\mathbf{l}_0)$ and $\hat{g}(a_0^{(1)})$.  \\
Step 3b: & Calculate the weights $w_0(a_0^{(1)},c)$. If $\hat{g}(a_0^{(1)})\leq c$, then ${m}_{w,t}(\bar{a}_t^{(1)},c)$ is undefined. \\
Step 4: & Estimate $\mathbb{E}(\tilde{Y}_0|{A}_0, \mathbf{L}_0)$. \\
Step 5: & Calculate $\hat{m}_{w,t}(\bar{a}_t^{(1)},c) = \hat{\mathbb{E}}_{w(c)}(Y_t^{\bar{a}_t^{(1)}}) = (\sum_{i=1}^n w_{0,i})^{-1}(\hat{w}_0(a_0^{(1)},c)^{\text{T}} \tilde{Y}_{-1})  $; that is, obtain the estimate of estimand 2 at $\bar{a}_t^{(1)}$ through calculating the weighted mean of the iterated outcome at $t=0$ under the respective intervention $a_0^{(1)}$. \\
\multicolumn{2}{l}{{\underline{\textbf{Then:}}}}  \\
Step 6: & Repeat steps 2-5 for the other interventions $\bar{A}_t^{(j)}$, $j=2,\ldots,n_a$. This yields an estimate of the desired dose-response curve (\ref{eqn:estimand_general}) at $t$. \\
Step 7: & Repeat steps 1-6 on $B$ bootstrap samples to obtain confidence intervals. \\
\hline
\end{tabular}
$^1$\tiny{The conditional treatment densities can be estimated with i) parametric models, if appropriate, like the linear model, ii) non-parametric flexible estimators, like highly-adaptive LASSO density estimation \cite{Hejazi:2022b}, iii) a ``binning strategy'' where a logistic regression model models the probability of approximately observing the intervention of interest at time $t$, given one has followed the strategy so far and given the covariates, iv) other options, like transformation models or generalized additive models of location, shape and scale \cite{Hothorn:2014, Stasinopoulos:2017}. Items i), ii), iii) are implemented in our package mentioned below.} \\
$^2$\tiny{The iterated weighted outcome regressions are recommended to be estimated data-adaptively, because the weighted outcomes are often non-symmetric. We recommend super learning for it \cite{vanderLaan:2011}, and this is what is implemented in the package mentioned below.}
\end{table}

\subsubsection{Multiple Interventions and Censoring}

So far, we have described the case of 1 intervention variable $A_t$ per time point. Of course, it is possible to employ the proposed methodology for multiple intervention variables $\mathbf{A}_t =(A_{t,1},A_{t,2},\ldots,A_{t,p})$. In this case, one simply has to estimate the conditional densities for all those intervention variables, and taking their product, at each time point. This follows from replacing (\ref{eqn:w1_long}) with ${g_s(\mathbf{a}_{s+1}\mid \mathbf h_{s+1})}/{g_s(\mathbf{a}_{s+1}\mid \mathbf{a}_s, \mathbf h_s)}$, and then factorizing the joint intervention distribution based on the assumed time ordering of $(L_s, A_{s,1}, A_{s,2},$ $\ldots, A_{s,p}, Y_s)$. Intervention variables may include censoring variables: we may, for instance, construct estimands under the intervention of no censoring. A possible option to apply the proposed methodology is then to use weights to target $E(Y_t^{\bar{C}_t=0}|A_t=a_t,\ldots,A_0=a_0)$ for units that do not have enough conditional support for the intervention strategy of interest. Appendix \ref{sec:appendix_theory_survival} lists the required modifications of the estimation procedure in this case. Alternatively, one may use weights that lead to $E(Y_t|C_t=c_t,A_t=a_t,\ldots C_0=c_0,A_0=a_0)$ under positivity violations. For this, one would require estimating the conditional censoring mechanisms in steps 1a and 1b too. There are however many subtleties and dangers related to the interpretation and estimation under censoring, and possibly competing events; for example, conditioning on the censoring indicators may lead to collider bias, intervening on censoring mechanisms to direct effect estimands \cite{Young:2020}, identification assumptions need to be refined and generic time-orderings where treatment variables are separated by different blocks of confounders may lead to more iterated expectations that have to be fitted. Those details go beyond the scope of this paper.

\subsection{Violations of the Positivity Assumption, Diagnostics and the Choice of $c$}\label{sec:framework_positivity}
Suppose we are interested in a subset of interventions $\bar{\mathcal{A}}_t^{\ast}$ that are part of the support region $\{\mathcal{A}_1 \times \ldots \times \mathcal{A}_t\}$. A violation of the strong positivity assumption exists if for any particular intervention of interest $(\tilde{a}_0,\ldots,\tilde{a}_t) \in \bar{\mathcal{A}}_t^{\ast}$ it happens that $g(\tilde{a}_t \mid \mathbf{h}_t) = 0$. To diagnose the extent of \textit{practical} positivity violations, one needs estimates of the conditional treatment densities, at each time point, and evaluated for each treatment strategy that is part of $\bar{\mathcal{A}}_t^{\ast}$ -- and summarize them in a meaningful way. We propose two options for facilitating this:

\begin{enumerate}
\item Calculate the proportion of weights (\ref{eqn:w_1t}) that are different from 1 (and thus indicate positivity violations) for each time point, for a range of $c$'s and the interventions of interest. Figure \ref{figure:4c} gives an example on how to visualize those proportions: it shows that for intervention trajectories close to $(0,\ldots,0)$ a high proportion of children have weights $\neq 1$ and thus for these interventions practical positivity violations exist (which we may want to address with the weighted estimand).
\item Estimate the conditional treatment densities, under the intervention trajectories of interest, with the following  \textit{binning strategy}: suppose $\bar{A}^{(j)}_t = (a_{0}^{(j)},\ldots,a_{t}^{(j)})$ is the $j^{\text{th}}$ element of $\bar{\mathcal{A}}_t^{\ast}$ and the intervention values at $t$ are already ordered such that $a_t^{(1)} < a_t^{(2)} <\ldots < a_t^{(n_a)}$; then calculate
    \begin{eqnarray}\label{eqn:calculate support}
&&P\left(a_t^{(j)} \in \left[a_t^{(j)} - \frac{1}{2}(a_t^{(j)}-a_t^{(j-1)}); \, a_t^{(j)} + \frac{1}{2}(a_t^{(j)}-a_t^{(j+1)}\right) \,\middle\vert\,  \right. \nonumber\\ && \left. \quad\quad\quad\quad\quad \bar{a}_{t-1}^{(j)} \in \left[a_{t-1}^{(j)} - \frac{1}{2}(a_{t-1}^{(j)}-a_{t-1}^{(j-1)}); \, a_{t-1}^{(j)} + \frac{1}{2}(a_{t-1}^{(j)}-a_{t-1}^{(j+1)}\right) ,\bar{l}_t\right)\,;
\end{eqnarray}
that is, we want to estimate the probability to approximately observe the intervention value $a_t^{(j)}$ under strategy $j$, given that one has followed the same strategy of interest $j$ so far (until $t-1$), and irrespective of the covariate history. Alternatively, instead of defining the bin widths through the intervention values of interest, we may calculate them data-adaptively \cite{Diaz:2011}. Estimating the mean of those probabilites over all observed $\bar{l}_t$ in a particular data set serves as diagnostic tool to measure the support for each rule $j$, at each time point. One could estimate (\ref{eqn:calculate support}) with standard regression techniques among the subset of those units who followed the respective strategy until $t-1$, and present a summary of those probabilites as a rough measure of support for each intervention trajectory of interest. This approach gives a sense of actual units following the strategies of interest in the data, given the covariates. Examples are given in Figure \ref{figure:results_extra1}.
\end{enumerate}

\textbf{Choice of c}: The first diagnostic, possibly visualized as in Figure \ref{figure:4c}, can be used to get a sense which set of $c's$ are informative to present weighted curves (as argued in Section \ref{sec:framework_both_estimands} below). More specifically, those $c's$ which are close enough to zero to detect positivity violations, but do not alter the intervention values of interest for a too large proportion of units (and for too many intervention values), are good candidates. Alternatively, as indicated above, similar to the case when truncating the propensity score for inverse probability´of treatment weighting or TMLE, ad-hoc choices could be used to decide when a treatment density is considered to be bounded away from zero (e.g., $0.01$).

\subsection{The Case for Presenting Both Estimands}\label{sec:framework_both_estimands}

Consider again our running example where we are interested in the probability of failure at 84 weeks under concentration levels of $(0,0,\ldots,0), \ldots, (6,6, \ldots, 6)$. As indicated above, adherent patients who are ultraslow metabolizers may not be able to have low concentration levels, e.g.  $g(0 \mid \text{ultraslow, adherent}) = 0$. In such a case, there is a strong case to redefine the estimand and not enforce to calculate $E(Y_t^{(0,0,\ldots,0)})$. Now, suppose we have $g_{84}(6 \mid \text{prior concentrations} = 6, \text{past covariates}) > 0$: obviously, there is no need to redefine the estimand. However, due to the finite sample the actual estimate may be $\hat{g}_{84}(6 \mid \text{prior concentrations} = 6, \text{past covariates}) \approx 0$. Obviously, we would in this region more likely rely on extrapolation than changing the estimand.

Now, practically, whenever an estimate of the conditional treatment density is close to zero, we do not necessarily know whether this is a finite sample (or estimation) issue, or due to ``infeasibility'', i.e. an illogical intervention given the history of a patient. Presenting estimand 1 shows the case where nothing is infeasible and we rely on extrapolations; presenting estimand 2, possibly shown for multiple $c's$, shows the change in the curve as we consider more intervention trajectories to be infeasible. Both estimands together give us a sense if main conclusions could change when positivity violations are addressed differently.

\section{Monte-Carlo Simulations}\label{sec:simulation}
In this section, we evaluate the proposed methods in three different simulation settings. Simulation 1 considers a very simple setting, as a basic reference for all approaches presented. In simulation 2, a survival setting is considered, to explore the stability of standard, unweighted g-computation in more sophisticated setups. The third simulation is complex, and inspired by the data-generating process of the data analysis. It serves as the most realistic setup for all method evaluations.

\subsubsection*{Data Generating Processes \& Estimands}
\noindent\textit{Simulation 1:} We simulated both a binary and normally distributed confounder, a continuous (normally distributed) intervention and a normally distributed outcome --  for 3 time points and a sample size of $n=10.000$. The exact model specifications are given in Appendix \ref{sec:appendix_DGP_1}. The intervention strategies of interest comprised intervention values in the interval $[2,11]$ which were constant over time; that is, $\bar{\mathcal{A}}_t^1 = \{(2,2,2),\ldots,(11,11,11)\}$ . The primary estimand $\psi^1$ is the CDRC (\ref{eqn:estimand_general}) for $t=1,2,3$ and all $\bar{a}_t \in \bar{\mathcal{A}}^1_t$. The secondary estimand $\psi^1_w$ is the weighted curve defined through (\ref{eqn:estimand2_time_multiple}), with the weights (\ref{eqn:w_1t}), for the same intervention strategies.\\

\noindent\textit{Simulation 2:} We simulated a continuous (normally distributed) intervention, two covariates (one of which is a confounder), an event outcome and a censoring indicator  --  for 5 time points and varying sample sizes of $n_1=500, n_2=1.000, n_3 = 10.000$ and $n_4=50.000$. The exact model specifications are given in Appendix \ref{sec:appendix_DGP_2}. The intervention strategies of interest comprised intervention values in the interval $[-7,13]$ which were constant over time; that is, $\bar{\mathcal{A}}_t^2 = \{(-7,-7,-7,-7,-7),\ldots,(13,13,13,13,13)\}$. The estimand $\psi^2$ is the CDRC (\ref{eqn:estimand_general}) for $t=1,2,3,4,5$, all $\bar{a}_t \in \bar{\mathcal{A}}^2_t$ and under no censoring ($\bar{c}_t=(0,\ldots,0)$).\\

\noindent\textit{Simulation 3:} We simulated data inspired by the data generating process of the data example outlined in Sections \ref{sec:motivation} and \ref{sec:data_analysis} as well as Figure \ref{fig:DAG}. The continuous intervention refers to drug concentration (of evavirenz), modeled through a truncated normal distribution. The binary outcome of interest is viral failure. Time-varying confounders, affected by prior interventions, are weight and adherence. Other variables include co-morbidities and drug dose (time-varying) we well as sex, genotype, age and NRTI regimen. We considered 5 clinic visits and a sample size of $n=1.000$.  The exact model specifications are given in Appendix \ref{sec:appendix_DGP_3}. The intervention strategies of interest comprised concentration values in the interval $[0,10]$ which were constant over time; that is, $\bar{\mathcal{A}}_t^3 = \{(0,0,0,0,0),\ldots,(10,10,10,10,10)\}$. The primary estimand $\psi_3$ is the CDRC (\ref{eqn:estimand_general}) for $t=1,2,3,4,5$ and all $\bar{a}_t \in \bar{\mathcal{A}}^3_t$. The secondary estimand $\psi^3_w$ is the weighted curve (\ref{eqn:estimand2_time_multiple}) with the weights (\ref{eqn:w_1t}) for the same intervention strategies.\\

\subsubsection*{Estimation \& Evaluation} All simulations were evaluated based on the results of $\mathcal{R}=1.000$ simulation runs. We evaluated the bias of estimating $\psi^1, \psi^2$ and $\psi^3$ with standard parametric g-computation with respect to the true CDRC. In simulations 1 and 3 model specification for $g$-computation was based on variable screening with LASSO, simulation 2 explored the idealized setting of using correct model specifications. We further evaluated the bias of the estimated weighted curve with the true weighted curve, for $c=1$; i.e. we looked whether $E(Y_t|A_t=a_t,\ldots,A_0=a_0)$ could be recovered for every intervention strategy of interest. Estimation was based on the algorithm of Table \ref{tab:estimation_gcomp_multiple}. The density estimates, which are required for estimating the weights, were based on both appropriate parametric regression models (that is, linear regression) and binning the continuous intervention into intervals; a computationally more sophisticated data-adaptive approach for density estimation is considered in Section \ref{sec:data_analysis}.  We then compared the estimated CDRC and weighted curve ($c=1$) with other weighted curves where $c=0.01$ and $c=0.001$. To estimate the iterated conditional expectations of the weighted outcome, we used super learning, i.e. a data adaptive approach \cite{vanderLaan:2011}. Our learner sets included different generalized linear (additive) regression models (with penalized splines, and optionally interactions), multivariate adaptive regression splines, LASSO estimates and regression trees -- after prior variable screening with LASSO and Cramer's V\cite{Heumann:2023}. Lastly, we visualize the conditional support of all considered intervention strategies, in all simulations, as suggested in (\ref{eqn:calculate support}).

\subsubsection*{Results}

\begin{figure}[ht!]
\begin{center}
\subfloat[Simulation 1: true and estimated CDRC at $t=2$]{\label{figure:2a}\includegraphics[scale=0.375]{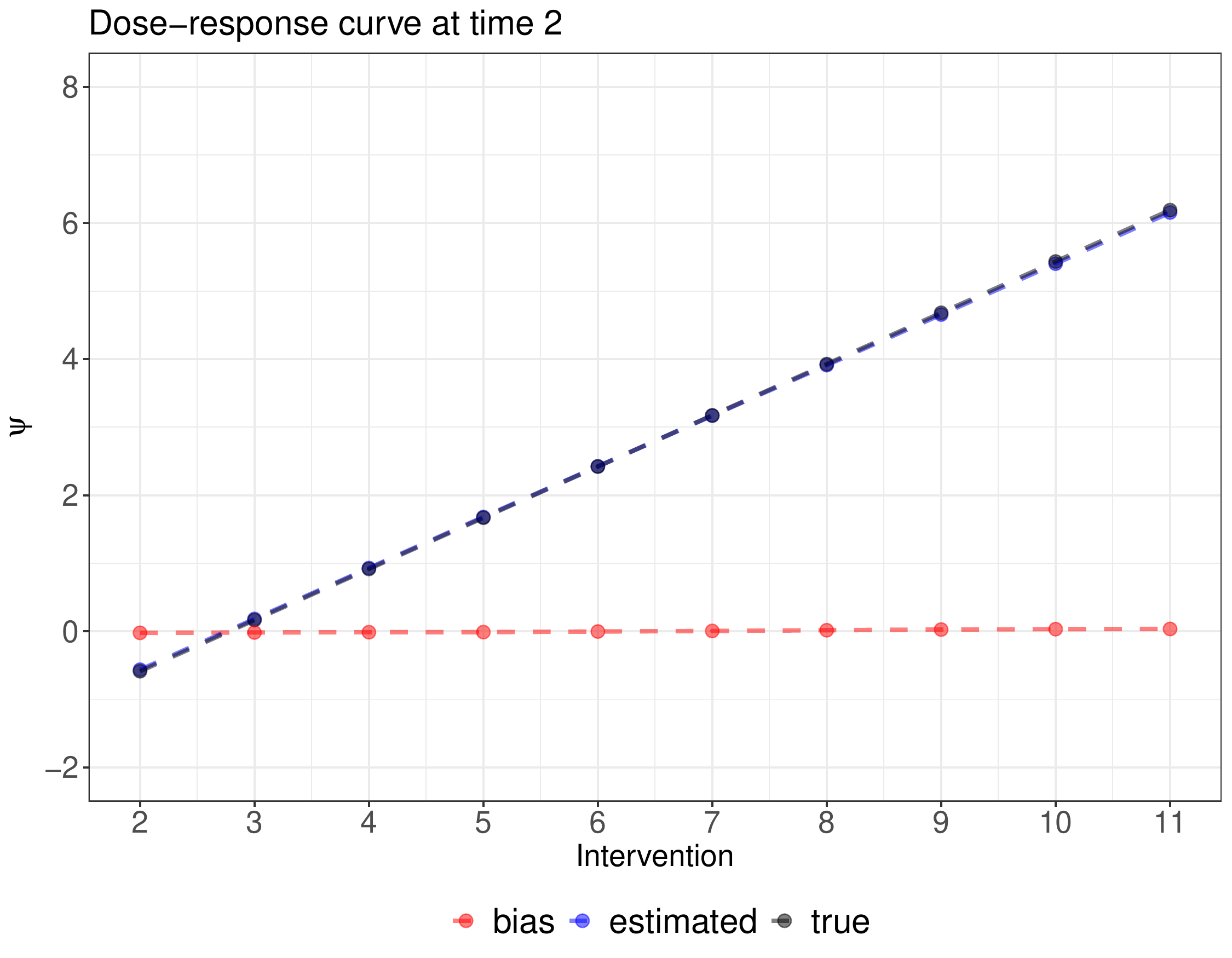}\hfill}
\subfloat[Simulation 1: weighted estimand, $c=1$, $t=1$]{\label{figure:2b}\includegraphics[scale=0.375]{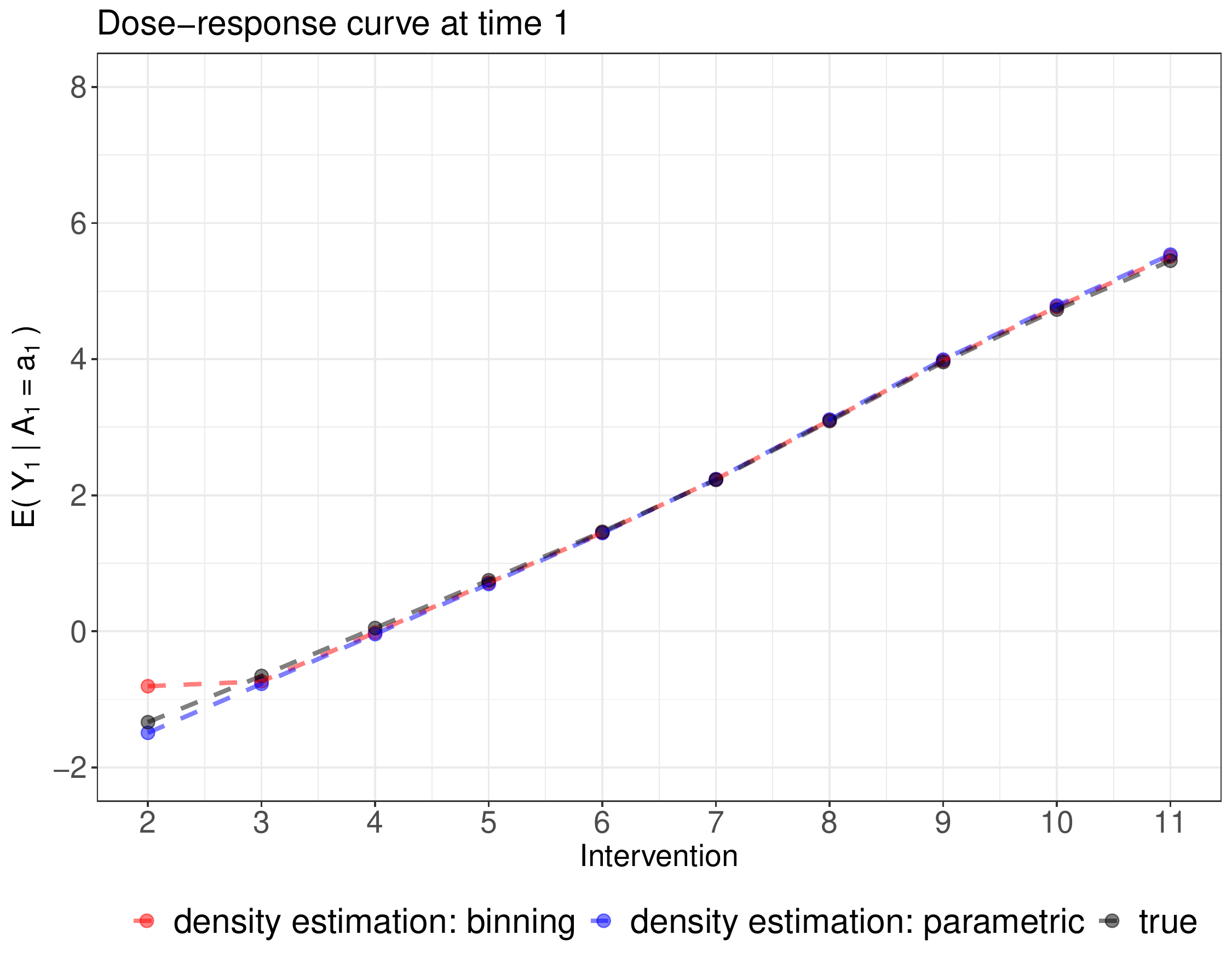}} \\
\subfloat[Simulation 2: true and estimated CDRC at $t=5$]{\label{figure:2c}\includegraphics[scale=0.375]{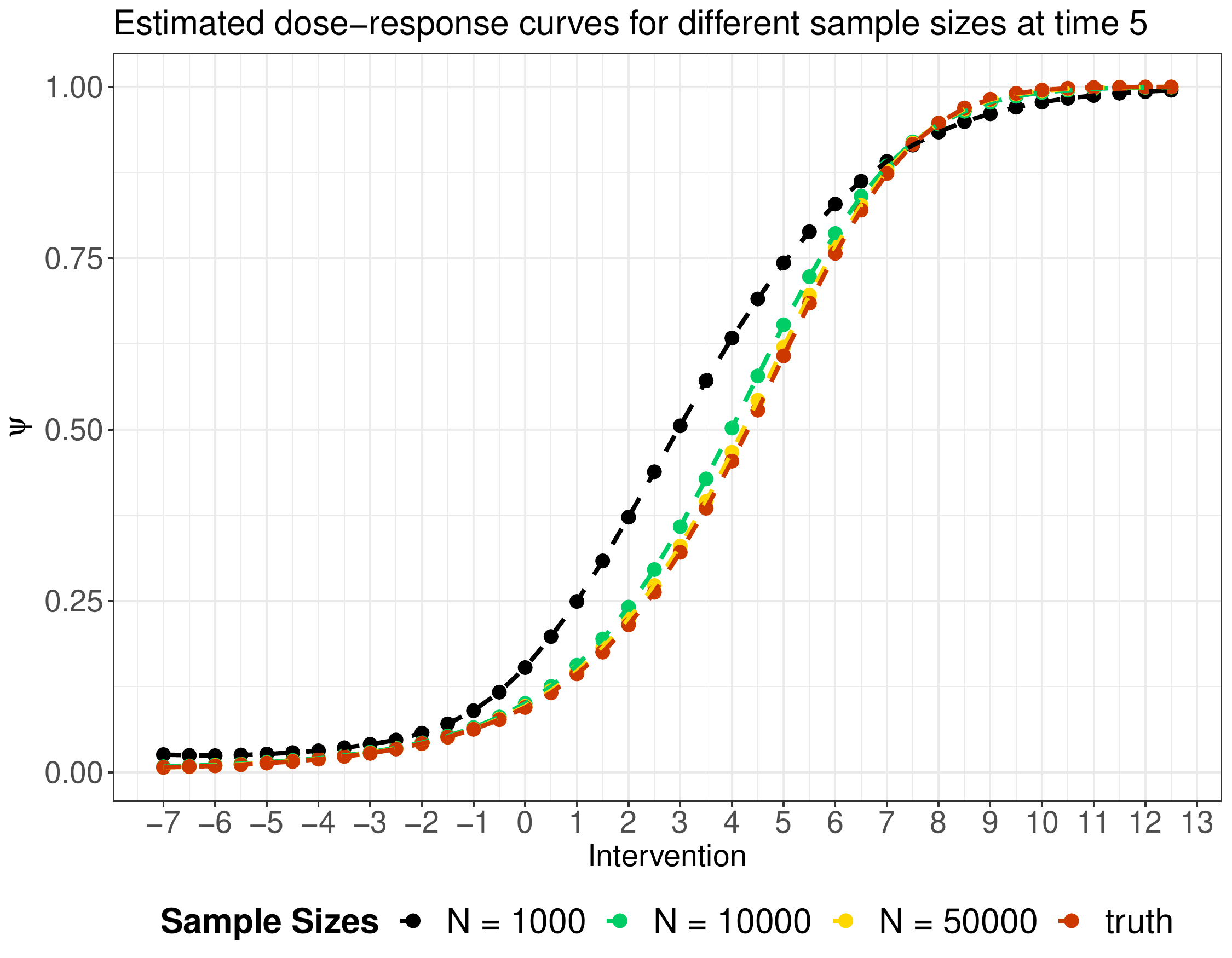}\hfill}
\subfloat[Simulation 3: true and estimated CDRC at $t=5$]{\label{figure:2d}\includegraphics[scale=0.375]{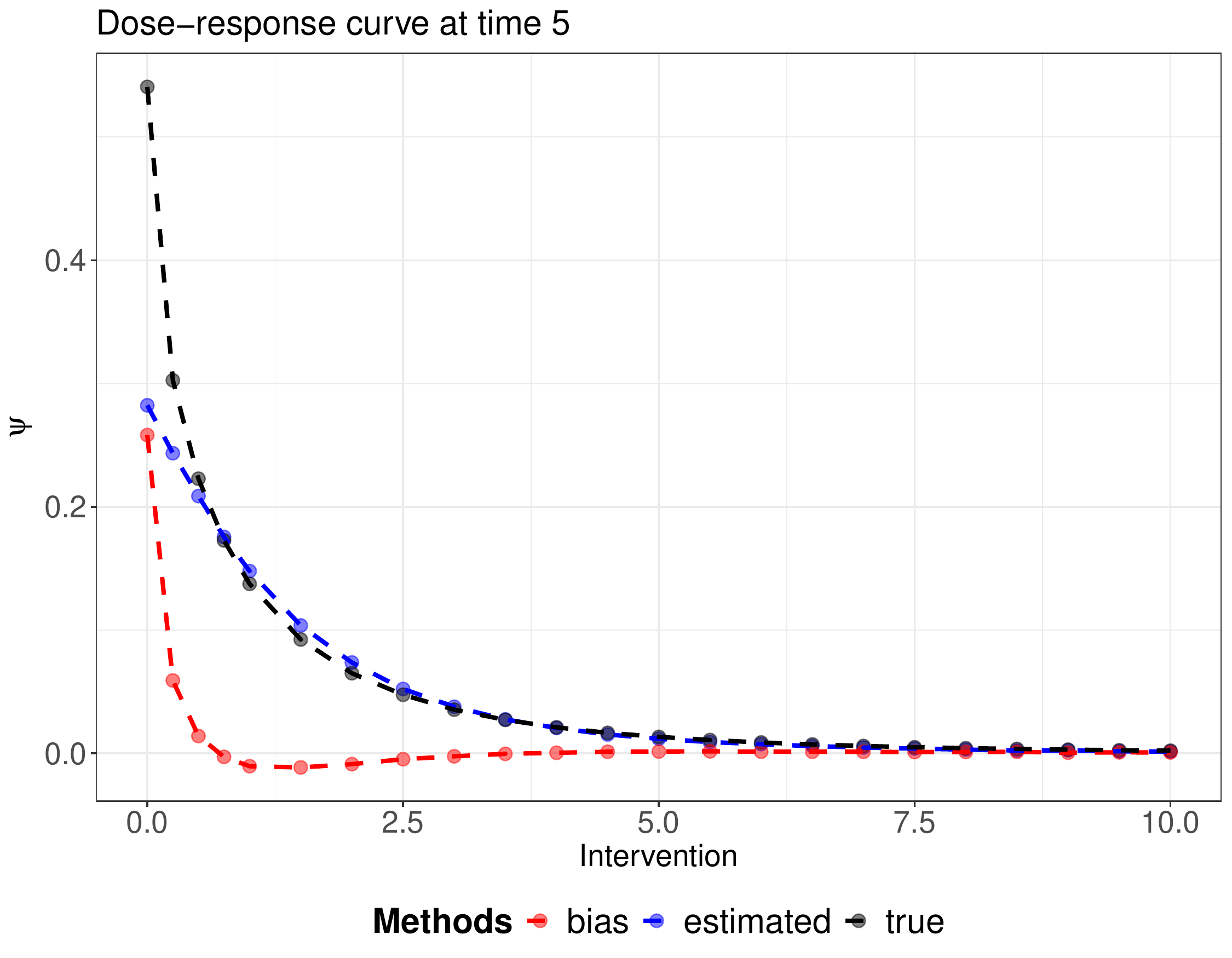}} \\
\subfloat[Simulation 3: weighted estimand, $c=1$, $t=1$]{\label{figure:2e}\includegraphics[scale=0.375]{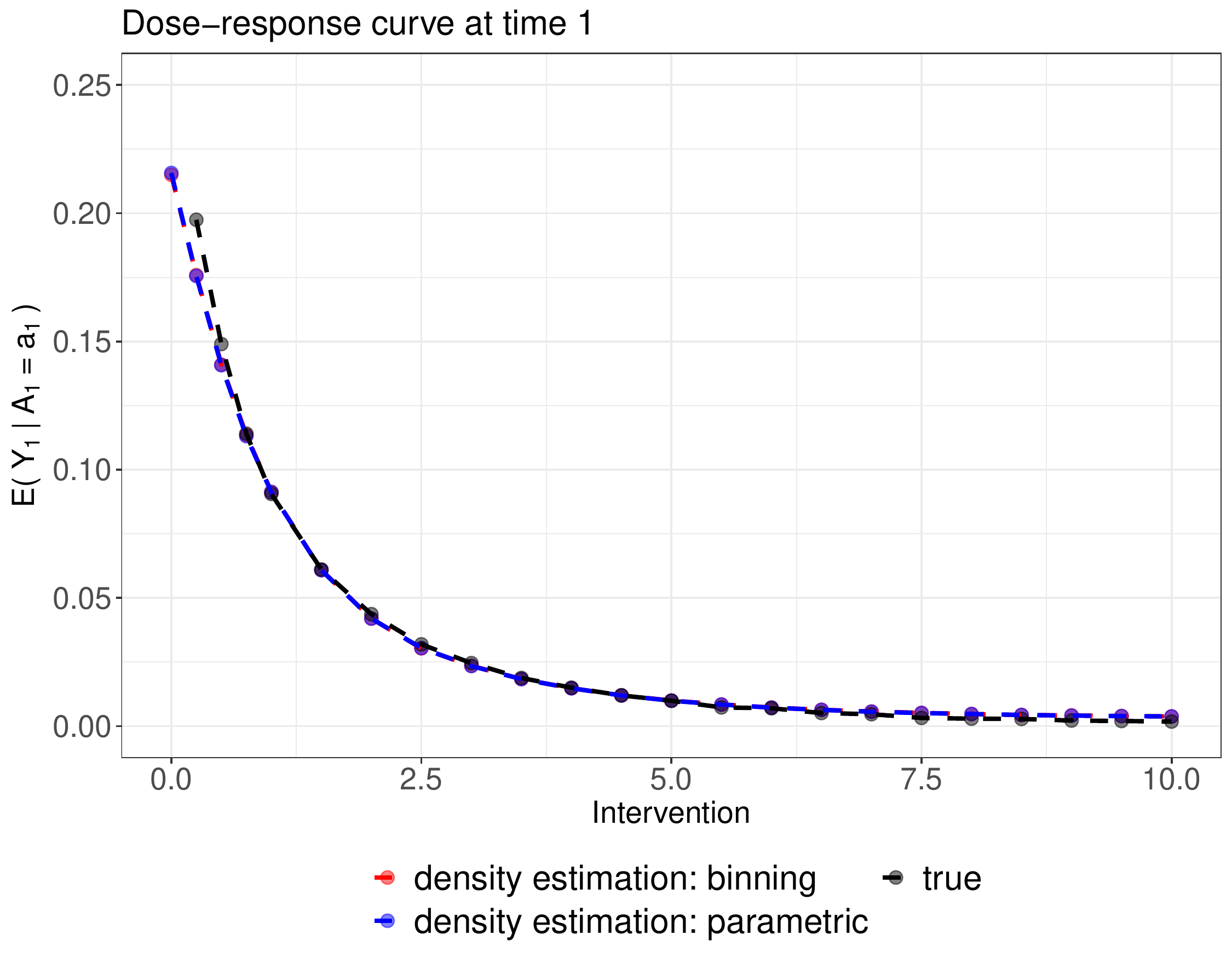}\hfill}
\subfloat[Simulation 3: CDRC's, $t=5$, $c=0.001, 0.01, 1$]{\label{figure:2f}\includegraphics[scale=0.375]{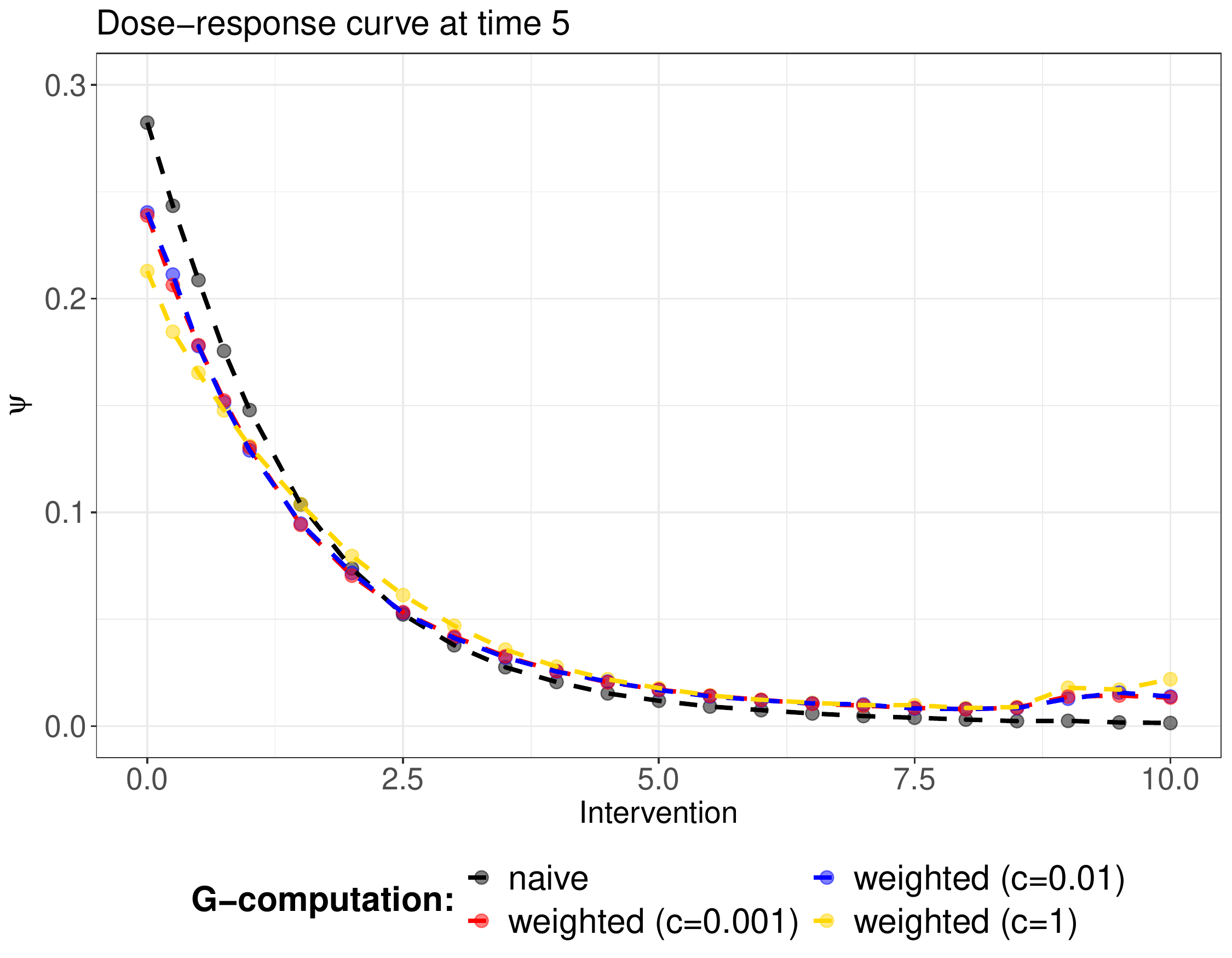}} \\
\caption{Results of the Monte-Carlo simulations}\label{figure:simulations}
\end{center}
\end{figure}

The results of the simulations are summarized in Figures \ref{figure:simulations}, \ref{figure:results_extra2} and \ref{figure:results_extra1}.

In simulation 1, applying standard g-computation to the continuous intervention led to approximately unbiased estimates for all intervention values, through all time points (Figure \ref{figure:2a}); and independent of the level of support (Figure \ref{figure:sim_support1}). The weighted curve, with $c=1$, recovered the association $E(Y_1|A_1=a_1)$ perfectly for the first time point, independent of the density estimation strategy (Figure \ref{figure:2b}). For later time points, some bias can be observed in regions of lower support (i.e., for intervention values between $9$ and $11$), see Figure \ref{figure:sim_extra1}.

Simulation 2 shows a setting in which g-computation estimates for the CDRC are not always unbiased, despite correct model specifications! Figure \ref{figure:2c} visualizes this finding for the fifth time point: it can be seen that a sample size of $\geq$ 10.000 is needed for approximately unbiased estimation of the CDRC over the range of all intervention strategies. Note, however, that for some time points lower sample sizes are sufficient to guarantee unbiased estimation (Figure \ref{figure:sim_extra2}).

The most sophisticated simulation setting 3 reveals some more features of the proposed methods. First, we can see that in areas of lowest support, towards concentration values close to zero (Figure \ref{figure:sim_support3}), there is relevant bias of a standard g-computation analysis (Figure \ref{figure:2d}); whereas in areas of reasonable to good support the CDRC estimates are approximately unbiased. The weighted curve with $c=1$ can recover the associations $E(Y_t|A_t=a_t,\ldots,A_1=a_1)$ (Figure \ref{figure:2e}) but there is some bias for later time points as the most critical area close to zero is approached, independent of the method used for estimating the weights (Figure \ref{figure:sim_extra4}). Both Figure \ref{figure:2f} and Figure \ref{figure:sim_extra5} highlight the behaviour of estimand 2 for $c=0.01$ and $c=0.001$: it can be clearly seen that the curves represent a compromise between the CDRC and the association represented by the weighted curve (with $c=1$). Evaluating the results for intervention values of zero, shows that weighting the curve in areas of low support yields to a compromise that moves the CDRC away from the estimated high probabilities of viral failure to more moderate values informed by the observed association. Knowledge of this behaviour may be informative for the data analysis below.

\section{Data Analysis}\label{sec:data_analysis}
We now illustrate the ideas based on the data from Bienczak et al. \cite{Bienczak:2016}, introduced in Section \ref{sec:motivation}. We consider the trial visits at $t=0,6,36,48,60,84$ weeks of $125$ children on an efavirenz-based treatment regime. The intervention of interest is efavirenz mid-dose interval concentration ($A_t$), defined as plasma concentration (in $mg/L$) 12 hours after dose; the outcome is viral failure ($Y_t$, defined as $> 100$ copies/mL). Measured baseline variables, which we included in the analysis, are $\mathbf{L_0}=\{\text{sex, genotype, age, the nucleoside reverse}$ $\text{transcriptase}$ $\text{inhib\-itors drug (NRTI), weight}\}$. Genoytpe refers to the metabolism status (slow, intermediate, extensive) related to the single nucleotide polymorphisms in the CYP2B6 gene, which is relevant for metabolizing evafirenz and directly affects its concentration in the body. Measured follow-up variables are $\mathbf{L}_t=\{\text{weight, adherence (measured through memory caps, MEMS), dose}\}$.

The assumed data generating process is visualized in the DAG in Figure \ref{fig:DAG}, and explained in more detail in Appendix \ref{sec:appendix_DAG}. Briefly, both weight and adherence are time-dependent confounders, potentially affected by prior concentration trajectories, which are needed for identification of the CDRC.

\begin{figure}[h!]
\begin{center}
{\noindent
\scalebox{1}{
\begin{tikzpicture}[x=6.75in,y=3.75in]
\node[color=black!60] (v0) at (0.120,-0.540) {Co-M$_0$};
\node[color=black!60] (v1) at (0.400,-0.540) {Co-M$_{36}$};
\node[color=black!60] (v2) at (0.550,-0.540) {Co-M$_{48}$};
\node[color=black!60] (v3) at (0.250,-0.540) {Co-M$_6$};
\node[color=black!60] (v4) at (0.700,-0.540) {Co-M$_{60}$};
\node[color=black!60] (v5) at (0.850,-0.540) {Co-M$_{84}$};
\node (v6) at (0.0100,-0.445) {Age};
\node (v11) at (0.140,-0.400) {Dose$_0$};
\node (v12) at (0.420,-0.400) {Dose$_{36}$};
\node[color=black!60] (v13) at (0.570,-0.400) {Dose$_{48}$};
\node (v14) at (0.270,-0.400) {Dose$_6$};
\node (v15) at (0.720,-0.400) {Dose$_{60}$};
\node (v16) at (0.870,-0.400) {Dose$_{84}$};
\node[color=black!60] (v17) at (0.160,-0.260) {EFV$_0$};
\node[color=green!60!black] (v18) at (0.440,-0.260) {EFV$_{36}$};
\node[color=black!60] (v19) at (0.590,-0.260) {EFV$_{48}$};
\node[color=green!60!black] (v20) at (0.290,-0.260) {EFV$_6$};
\node[color=green!60!black] (v21) at (0.740,-0.260) {EFV$_{60}$};
\node[color=green!60!black] (v22) at (0.890,-0.260) {EFV$_{84}$};
\node (v23) at (0.0600,-0.368) {Genotype};
\node (v24) at (0.360,-0.800) {MEMS$_{36}$};
\node (v25) at (0.510,-0.800) {MEMS$_{48}$};
\node[color=black!60] (v26) at (0.210,-0.800) {MEMS$_6$};
\node (v27) at (0.660,-0.800) {MEMS$_{60}$};
\node[color=black!60] (v28) at (0.810,-0.800) {MEMS$_{84}$};
\node (v29) at (0.0800,-0.800) {NRTI$_0$};
\node (v30) at (0.000,-0.602) {Sex};
\node (v31) at (0.180,-0.0700) {VL$_0$};
\node (v32) at (0.460,-0.0700) {VL$_{36}$};
\node (v33) at (0.610,-0.0700) {VL$_{48}$};
\node[color=black!60] (v34) at (0.310,-0.0700) {VL$_6$};
\node (v35) at (0.760,-0.0700) {VL$_{60}$};
\node[color=blue!80!black] (v36) at (0.910,-0.0700) {VL$_{84}$};
\node (v38) at (0.100,-0.680) {Weight$_0$};
\node (v39) at (0.380,-0.680) {Weight$_{36}$};
\node[color=black!60] (v40) at (0.530,-0.680) {Weight$_{48}$};
\node (v41) at (0.230,-0.680) {Weight$_6$};
\node (v42) at (0.680,-0.680) {Weight$_{60}$};
\node (v43) at (0.830,-0.680) {Weight$_{84}$};
\draw [->] (v0) edge (v3);
\draw [->] (v0) edge (v34);
\draw [->] (v0) edge (v41);
\draw [->] (v1) edge (v2);
\draw [->] (v1) edge (v33);
\draw [->] (v1) edge (v40);
\draw [->] (v2) edge (v4);
\draw [->] (v2) edge (v35);
\draw [->] (v2) edge (v42);
\draw [->] (v3) edge (v1);
\draw [->] (v3) edge (v32);
\draw [->] (v3) edge (v39);
\draw [->] (v4) edge (v5);
\draw [->] (v4) edge (v36);
\draw [->] (v4) edge (v43);
\draw [->] (v6) edge (v0);
\draw [->] (v6) to [out=-5,in=135, looseness=0.1] (v1);
\draw [->] (v6) to [out=-5,in=135, looseness=0.1] (v2);
\draw [->] (v6) to  (v3);
\draw [->] (v6) to [out=-5,in=135, looseness=0.1]  (v4);
\draw [->] (v6) to [out=-5,in=135, looseness=0.15]  (v5);
\draw [->] (v6) edge (v29);
\draw [->] (v6) edge (v38);
\draw [->] (v11) edge (v14);
\draw[->] (v11) to  [bend right, looseness=0.5] (v17);
\draw [->] (v12) edge (v13);
\draw[->] (v12) to  [bend right, looseness=0.5] (v18);
\draw [->] (v13) edge (v15);
\draw[->] (v13) to  [bend right, looseness=0.5] (v19);
\draw [->] (v14) edge (v12);
\draw[->] (v14) to  [bend right, looseness=0.5] (v20);
\draw [->] (v15) edge (v16);
\draw[->] (v15) to  [bend right, looseness=0.5] (v21);
\draw[->] (v16) to  [bend right, looseness=0.5] (v22);
\draw [->] (v17) edge (v31);
\draw [->] (v17) edge (v32);
\draw [->] (v17) edge (v33);
\draw [->] (v17) edge (v34);
\draw [->] (v17) edge (v35);
\draw [->] (v17) edge (v36);
\draw [->] (v18) edge (v32);
\draw [->] (v18) edge (v33);
\draw [->] (v18) edge (v35);
\draw [->] (v18) edge (v36);
\draw [->] (v19) edge (v33);
\draw [->] (v19) edge (v35);
\draw [->] (v19) edge (v36);
\draw [->] (v20) edge (v32);
\draw [->] (v20) edge (v33);
\draw [->] (v20) edge (v34);
\draw [->] (v20) edge (v35);
\draw [->] (v20) edge (v36);
\draw [->] (v21) edge (v35);
\draw [->] (v21) edge (v36);
\draw [->] (v22) edge (v36);
\draw [->] (v23) edge (v17);
\draw [->] (v23) to [out=5,in=230, looseness=0.15] (v18);
\draw [->] (v23) to [out=5,in=230, looseness=0.15] (v19);
\draw [->] (v23) to [out=5,in=230, looseness=0.15] (v20);
\draw [->] (v23) to [out=5,in=230, looseness=0.15] (v21);
\draw [->] (v23) to [out=5,in=230, looseness=0.15] (v22);
\draw [->] (v24) edge (v2);
\draw [->] (v24) edge (v18);
\draw [->] (v24) edge (v25);
\draw [->] (v25) edge (v4);
\draw [->] (v25) edge (v19);
\draw [->] (v25) edge (v27);
\draw [->] (v26) edge (v1);
\draw [->] (v26) edge (v20);
\draw [->] (v26) edge (v24);
\draw [->] (v27) edge (v5);
\draw [->] (v27) edge (v21);
\draw [->] (v27) edge (v28);
\draw [->] (v28) edge (v22);
\draw [->] (v30) edge (v0);
\draw [->] (v30) edge (v38);
\draw [->] (v34) edge (v1);
\draw [->] (v31) edge (v3);
\draw [->] (v31) edge (v34);
\draw [->] (v32) edge (v2);
\draw [->] (v32) edge (v33);
\draw [->] (v33) edge (v4);
\draw [->] (v33) edge (v35);
\draw [->] (v34) edge (v32);
\draw [->] (v35) edge (v5);
\draw [->] (v35) edge (v36);
\draw[->] (v38) to  [out=35,in=-35, looseness=0.5] (v11);
\draw [->] (v38) edge (v41);
\draw[->] (v39) to  [out=35,in=-35, looseness=0.5] (v12);
\draw [->] (v39) edge (v40);
\draw[->] (v40) to  [out=35,in=-35, looseness=0.5] (v13);
\draw [->] (v40) edge (v42);
\draw[->] (v41) to  [out=35,in=-35, looseness=0.5] (v14);
\draw [->] (v41) edge (v39);
\draw[->] (v42) to  [out=35,in=-35, looseness=0.5] (v15);
\draw [->] (v42) edge (v43);
\draw[->] (v43) to  [out=35,in=-35, looseness=0.5] (v16);
\draw [->] (v0) edge (v26);
\draw [->] (v3) edge (v24);
\draw [->] (v1) edge (v25);
\draw [->] (v2) edge (v27);
\draw [->] (v4) edge (v28);
\draw [->] (v38) edge (v3);
\draw [->] (v41) edge (v1);
\draw [->] (v39) edge (v2);
\draw [->] (v40) edge (v4);
\draw [->] (v42) edge (v5);
\draw [->] (v17) to [out=310,in=135, looseness=0.15] (v26);
\draw [->] (v20) to [out=310,in=135, looseness=0.15] (v24);
\draw [->] (v18) to [out=310,in=135, looseness=0.15] (v25);
\draw [->] (v19) to [out=310,in=135, looseness=0.15] (v27);
\draw [->] (v21) to [out=310,in=135, looseness=0.15] (v28);
\node  (b1)  at (-0.025,-0.9)  {};
\node  (b2)  at (0.95,-0.9)  {};
\draw[->] (b1) -- (b2) node[midway,below] {\footnotesize Time};
\end{tikzpicture}
}
}
\end{center}\caption{Directed Acyclic graph for the data analysis. The intervention variable is shown in green (efavirenz concentration), the outcome in blue (viral load at the end of follow-up). Unmeasured variables are colored in grey. Both MEMS and weight are time-dependent confounders which are affected by prior treatment nodes. }\label{fig:DAG}
\end{figure}
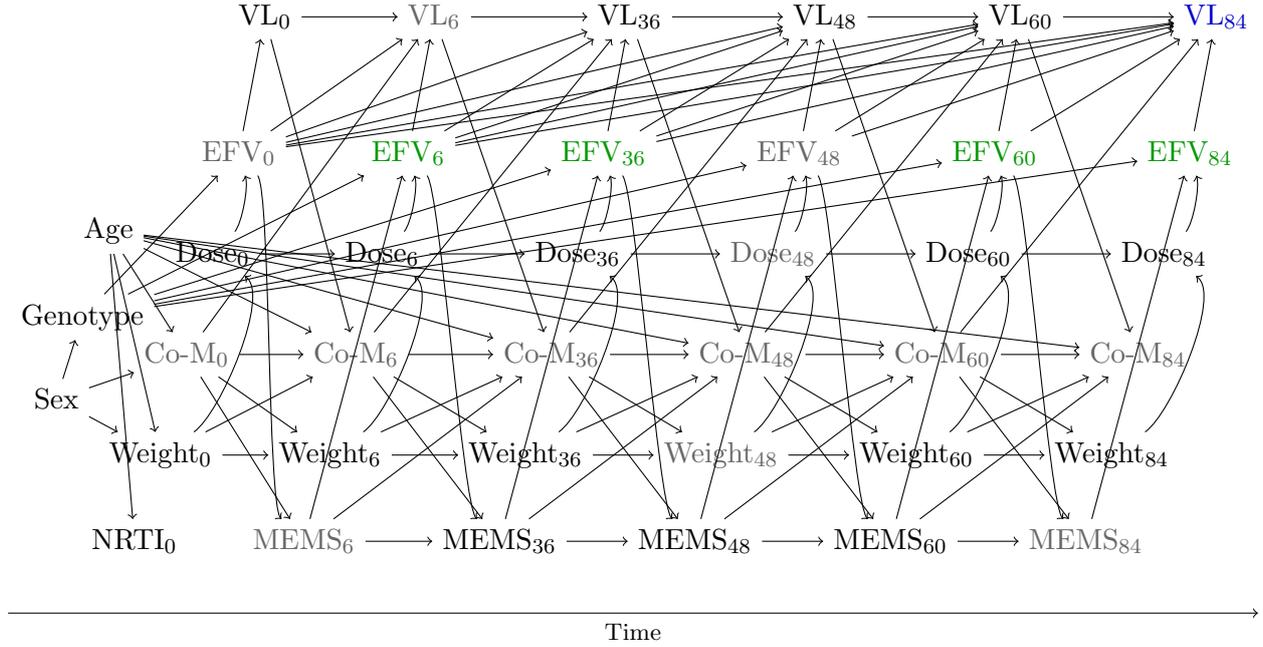

Our target estimands are the CDRC (\ref{eqn:estimand_general}) and the weighted curve (\ref{eqn:estimand2_time_multiple}) at $t=84$ weeks. The intervention strategies of interest are $\bar{\mathcal{A}}_t = \{(0,0,0,0),\ldots,(6,6,6,6)\}$.

The analysis illustrates the ideas based on a complete case analysis of all measured variables represented in the DAG ($n=58$), but excluding dose (not needed for identification) and MEMS (due to the high proportion of missingness) \cite{Holovchak:2024}.

We estimated the CDRC both with sequential and parametric g-computation using the intervention strategies $\bar{\mathcal{A}}_t$. The estimation of the weighted curve followed the algorithm of Table \ref{tab:estimation_gcomp_multiple}. The conditional treatment densities, which are needed for the construction of the weights, were estimated both parametrically based on the implied distributions from linear models (for concentrations under 5 $mg/L$) and  with highly-adaptive LASSO conditional density estimation (for concentrations $\geq 5$ $mg/L$) \cite{Hejazi:2022, Hejazi:2022b}. This is because the density in the lower concentration regions were approximately normally distributed, but more complex for higher values.

The conditional expectation of the weighted outcome (Step 4 of the algorithm) was estimated data-adaptively with super learning using the following learning algorithms: multivariate adaptive regression splines, generalized linear models (also with interactions), ordinal cumulative probability models, generalized additive models as well as the mean and median. Prior variable screening, which was essential given the small sample size, was based on both the LASSO and Cramer's $V$ .

We estimated the weighted curve for $c=0.001, 0.01, 0.025, 0.2, 1$. We also calculate the support of the continuous intervention strategies $\bar{\mathcal{A}}_t$ as described in Section \ref{sec:framework_positivity}.

\subsubsection*{Results}
\begin{figure}[h!]
\begin{center}
\subfloat[CDRC, weighted curve and conditional support, represented through background shading. Support is defined as proportion of weights, which are unequal to 1, see text for details.]{\label{figure:4a}\includegraphics[scale=0.425]{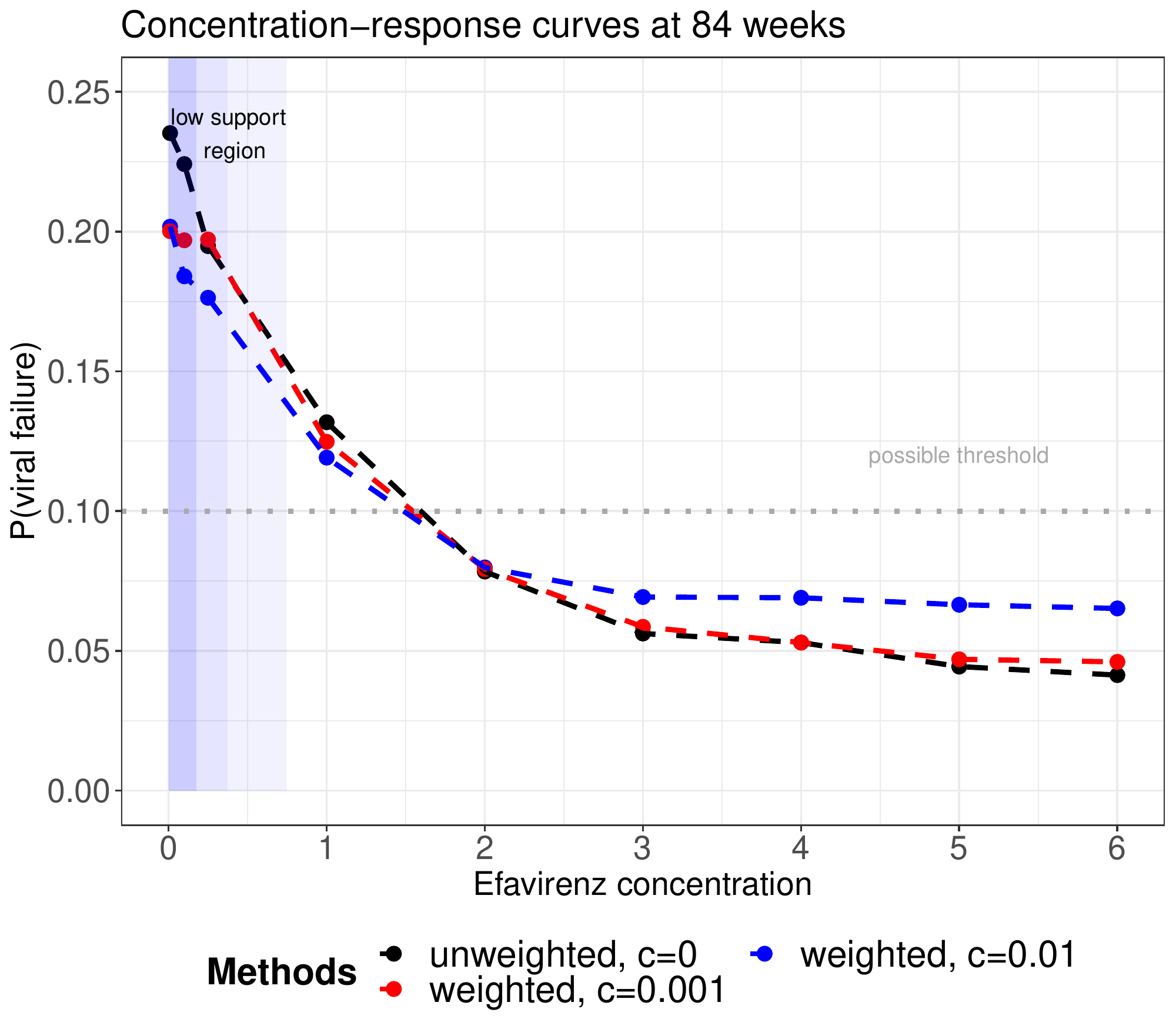}\hfill}
\subfloat[CDRC and weighted curves]{\label{figure:4b}\includegraphics[scale=0.425]{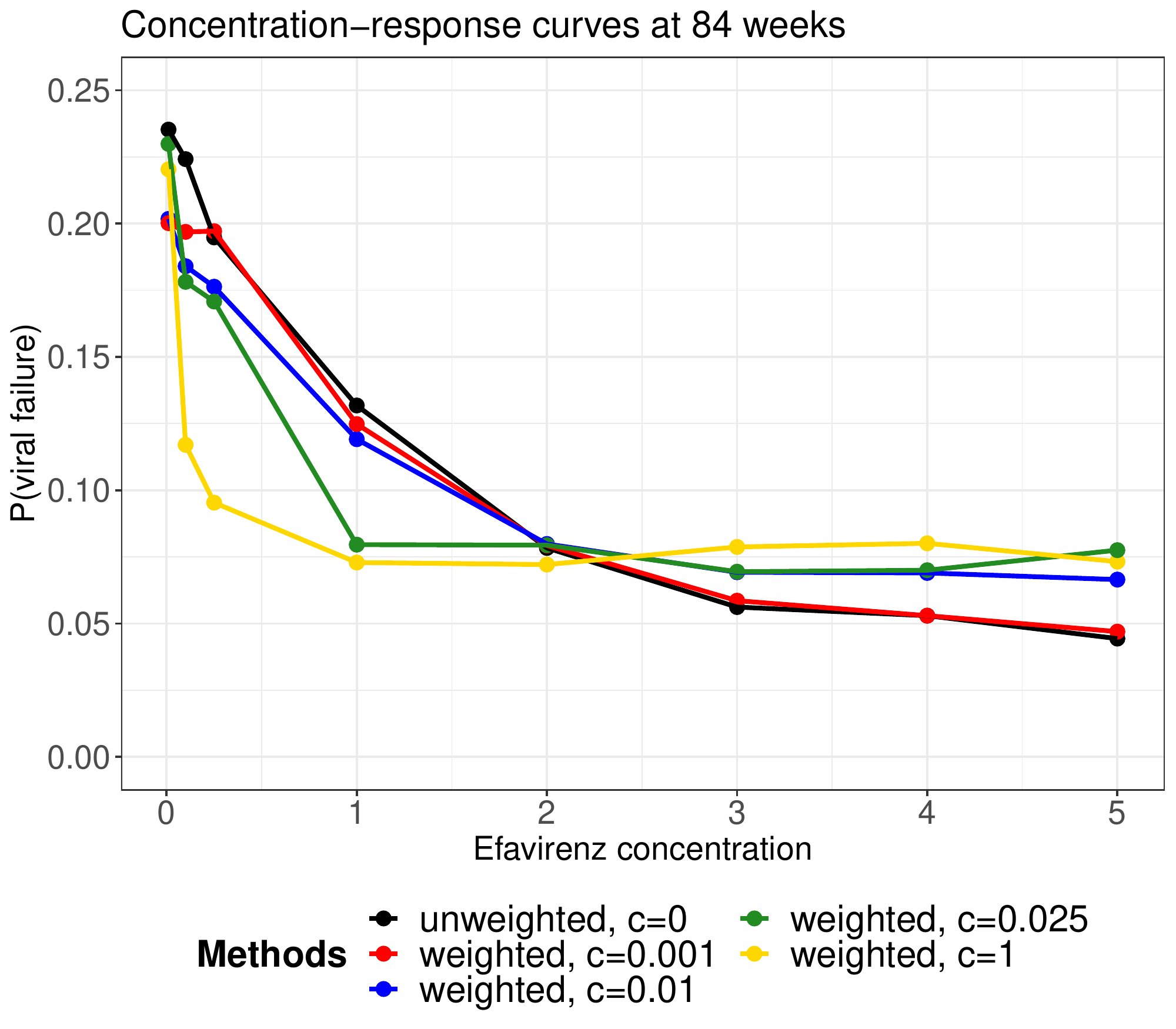}} \\
\subfloat[Conditional support and weight behaviour: percentage of weights which are unequal 1, as a function of $c$ and concentration values, averaged over all visits. ]{\label{figure:4c}\includegraphics[scale=0.4]{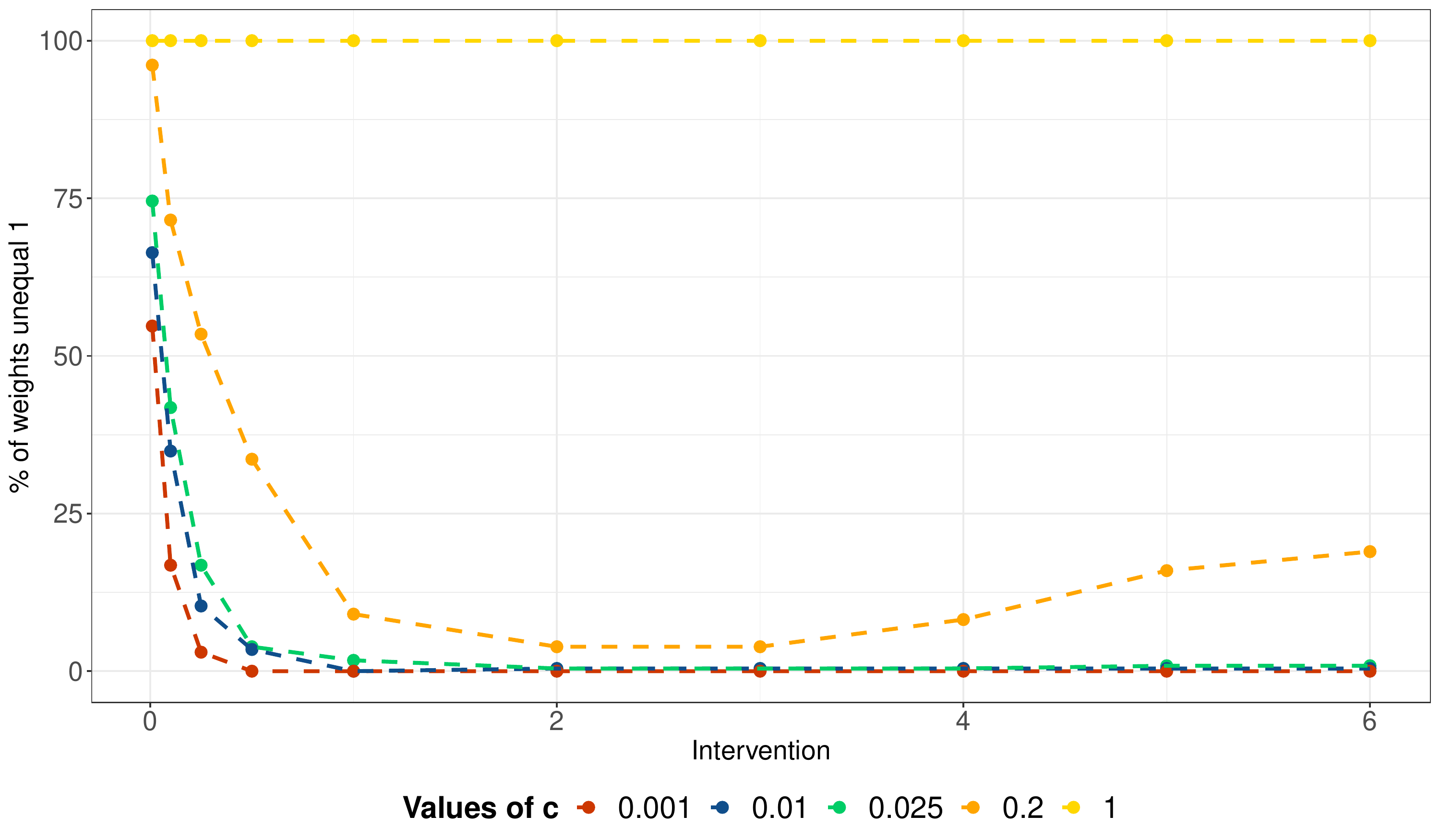}\hfill}
\\
\caption{Results of the data analysis}\label{figure:data_analysis}
\end{center}
\end{figure}

The main results of the analyses are given in Figure  \ref{figure:data_analysis}.

Figures \ref{figure:4a} and \ref{figure:4b} show that the estimated CDRC (black solid line) suggests higher probabilities of failure with lower concentration values.  The curve is steep in the region from $0-2$mg, which is even more pronounced for parametric g-computation when compared to sequential g-computation (Figure \ref{figure:data_extra6}).

Both Figures \ref{figure:4a} and \ref{figure:data_support4} illustrate the low level of (conditional support) support for extremely small concentration values close to $0$ mg/L.

Note that in Figure \ref{figure:4a} the area shaded in dark blue relates to low conditional support, defined as a proportion $>50$\% of weights being unequal from 1; that is, more than 50\% of observations having an estimated conditional treatment density $<0.01$ under the respective intervention values. Intermediate, light and no blue background shadings refer to percentages of $15-50$\%, $5-15$\% and $<5$\%, respectively. More details on how the conditional support defined through weight summaries varies as a function of $c$ and the intervention values is illustrated in Figure \ref{figure:4c}. It can be seen that with a choice of $c \leq 0.025$, for most patients the actual intervention of interest seems feasible for concentrations values above $0.5$ mg/l; however, concentrations $\leq 0.5$ mg/l seem unrealistic for many patients and thus weighting is then used more often. As the proportion of weights, which are unequal 1, do not change much between $c=0.01$ and $c=0.025$, we report only curves for $c=0.001$ and $c=0.01$ in Figure \ref{figure:4a}.

The weighted curves in Figure \ref{figure:4b} are less steep in the crucial areas close to zero and --as in the simulation studies-- we observe a tendency of the curves with $c=0.001, 0.01, 0.025$ to provide a visual compromise between the CDRC and the estimand recovering the relevant association ($c=1$).

The weighted curves show the probability of failure under different concentration levels, at least if the respective concentration level is likely possible (realistic, probable) to be observed given the patient's covariate history; for patients where this seems unlikely, patients rather have individual concentrations that lead to outcomes comparable to the patient population with the respective concentration trajectory. For example, if patients are adherent and slow metabolizers, it may be unlikely that they achieve low concentration levels and we do not enforce the intervention of interest. We let these patients rather behave individually in line with otucomes typically seen for low concentration values.

The practical implication is that if we do not want to extrapolate in regions of low support and/or do not consider specific concentration values to be feasible for some patients, the concentration-response curve is flatter. The current lower recommended target concentration limit is $1$ mg/l, at which we still estimate high failure probabilities under standard g-computation approaches.  \textit{If, for example, we do not want to accept failure probabilites $<10$\% --but are worried about positivity violations--, then we see that main conclusions with respect to a lower recommended target concentration limit would not change when using estimand 2 in addition to estimand 1 (Figure \ref{figure:4a}).}

We recommend to report results similar to Figure \ref{figure:4a}, that is presenting a figure that shows both estimand 1 and estimand 2 (possibly based on several $c's$, informed by a graph as in Figure \ref{figure:4c}).

\section{Discussion}\label{sec:discussion}
We have introduced and evaluated various methods for causal effect estimation with continuous multiple time-point interventions. As an obvious first step, we investigated standard (parametric or sequential) g-formula based approaches for identification and estimation of the actual causal dose-response curve. We emphasized that this has the advantage of sticking to the estimand of interest, but with a relatively strong positivity assumption which will likely be violated in most data analyses. Our simulations were designed to explore how well such a standard approach may extrapolate in sparse data regions and how the approach performs in both simple and complex settings. We found that in simple scenarios the CDRC estimates were approximately unbiased, but that in more complex settings sometimes large sample sizes were needed for good performance and that in regions of very poor support a relatively large bias could be observed. These findings suggest that the toolkit for causal effect estimation with longitudinal continuous interventions should therefore ideally be broader.

We therefore proposed the use of a weight function which returns the CDRC under enough conditional support, and makes use of the crude associations between the outcome at time $t$ and the treatment history otherwise. This choice ensures that the estimand is always well-defined and one does not require positivity. Our simulations suggest that the provided estimation algorithm can indeed succesfully recover the weighted estimand and illustrate the compromise that is practically achieved. As the proposed weights may have skewed distributions, we highlighted the importance of using a data-adaptive estimation approach, though appropriate interval estimation, beyond using bootstrapping, remains to be investigated.

We hope that our manuscript has shown that for continuous multiple time point interventions one has to ultimately make a tradeoff between estimating the CDRC as closely as possible, at the risk of bias due to positivity violations and minimizing the risk of bias due to positivity violations, at the cost of redefining the estimand. If the scientific question of interest allows a redefinition of the estimand in terms of longitudinal modified treatment policies \cite{Diaz:2020, Diaz:2024, Hoffman:2024} then this may be a great choice. Otherwise, it may be helpful to use g-computation algorithms for continuous interventions, visualize the estimates in appropriate curves, calculate the suggested diagnostics by estimating the conditional support and then estimate the proposed weighted curves as a magnifying glass for the curves behaviour in areas of low support (where the intervention of interest may not be feasible or realistic).

Our suggestions can be extended in several directions: it may be possible to design different weight functions that keep the spirit of modifying estimands under continuous interventions only in areas of low support; the extent of the compromise, controlled by the tuning parameter $c$, may be chosen data-adaptively and our basic considerations for time-to-event data may be extended to cover more estimands, particular in the presence of competing risks.

\subsubsection*{Software} All approaches considered in this paper have been implemented in the $R$-packages \texttt{CICI} and \texttt{CICIplus} available at \url{https://cran.r-project.org/web/packages/CICI/index.html} and \url{https://github.com/MichaelSchomaker/CICIplus}.

\subsubsection*{Acknowledgements}
We are grateful for the support of the CHAPAS-3 trial team, their advise regarding the illustrative data analysis and making their data available to us. We would like to particularly thank David Burger, Sarah Walker, Di Gibb, Andrzej Bienczak and Elizabeth Kaudha. We would also like to acknowledge Daniel Saggau, who has contributed to the data-generating processes of our simulation setups and Igor Stojkov for contributing substantive knowledge to pharmacological angles relevant to the data example.  Computations were performed using facilities provided by the University of Cape Town's ICTS High Performance Computing team. Michael Schomaker is supported by the German Research Foundations (DFG) Heisenberg Programm (grants 465412241 and 465412441).

\bibliographystyle{unsrtnat}
{\footnotesize
\bibliography{literature}
}


\clearpage
\appendix

\section{Additional Theory Details.}\label{sec:appendix_theory}

\subsection{Iterated Nested Expectation Representation of Estimand 1}\label{sec:appendix_theory_CDRC_details}
Using (\ref{eqn:CDRC_long_recursive}) corresponds to a recursive definition of estimand 1. For $s=t$, the integral corresponds to
\begin{eqnarray*}
{m}_t (a_t, \mathbf{h}_t) &=& E(Y_t \mid A_t=a_t,\mathbf{H}_t=\mathbf{h}_t)\,.
\end{eqnarray*}
Moving to $s=t-1$ yields
\begin{eqnarray*}
{m}_{t-1} (a_{t-1}, \mathbf{h}_{t-1})  &=& E({m}_{t} (a_t, h_t) \mid A_{t-1}=a_{t-1},\mathbf{H}_{t-1}=\mathbf{h}_{t-1}) \\
 &=& E(E(Y_t \mid \bar{A}_t = \bar{a_t}, \mathbf{H}_t=\mathbf{h}_t) \mid \bar{A}_{t-1} = \bar{a}_{t-1}, \mathbf{H}_{t-1}=\mathbf{h}_{t-1})\,.
\end{eqnarray*}
Reevaluating the integral from $s=t-2,\ldots,0$ then leads to
\begin{eqnarray*}
{m}_0 (a_0, \mathbf{h}_0)  &=& \mathbb{E}(\,\ldots\mathbb{E}(\,\mathbb{E}(Y_t|\bar{{A}}_{t}=\bar{a}_{t}, \mathbf{\bar{L}}_t) | \bar{{A}}_{t-1}=\bar{a}_{t-1}, \mathbf{\bar{L}}_{t-1}\,)\ldots|{A}_{0}={a}_{0}, \mathbf{{L}}_{0}\,)\,)\,.
\end{eqnarray*}
This expression is identical to $\mathbb{E}(Y_t^{\bar{a}_t})$ under conditional sequential exchangeability, positivity and consistency, which yields the sequential g-formula listed in (\ref{eqn:seq_g_formula}). The result is known from Bang and Robins \cite{Bang:2005}. Briefly, first recall that positivity can be defined as in (\ref{eqn:positivity_longitudinal}), consistency is the requirement that
\begin{eqnarray}
Y^{\bar{a}_t}_t &=& Y_t \text{ if } \bar{A}_{t} = \bar{a}_{t}  \,,
\end{eqnarray}
and  sequential conditional exchangeability is defined as
\begin{eqnarray}
Y^{\bar{a}_t}_t\coprod A_{t}|{H}_{t} \,.
\end{eqnarray}
Suppose $t=2$. Then,
\begin{eqnarray*}
E(Y_1^{a_1, a_0}) &=& E(E(Y_1^{a_1, a_0} \mid L_0 ))\\
&=& E(E(Y_1^{a_1, a_0} \mid  L_0, A_0=a_0 ))\\
&=& E(E(E(Y_1^{a_1, a_0} \mid A_0=a_0, L_0, L_1 )) \mid L_0, A_0=a_0 ) \\
&=& E(E(E(Y_1^{a_1, a_0} \mid A_0=a_0, L_0, L_1, A_1=a_1 )) \mid L_0, A_0=a_0 ) \\
&=& E(E(E(Y_1^{a_1, a_0} \mid \bar{A}_1=\bar{a}_1, \bar{L}_1 )) \mid L_0, A_0=a_0 ) \\
&=& E(E(E(Y_1 \mid \bar{L}_1, \bar{A}_1=\bar{a}_1 ))\mid L_0, A_0=a_0 )\\
\end{eqnarray*}
where the first and third equality follow from the law of iterated expectation, the second and fourth by conditional exchangeability (i.e., $Y^{a_1,a_0} \coprod A_0 \mid L_0$ and $Y^{a_1,a_0} \coprod A_1 \mid L_0, L_1, A_0$), the fifth by definition and the sixth by consistency. Similarly, we can derive (\ref{eqn:seq_g_formula}) generically for any $t \in \mathcal{T}$. For the conditional expectations to be well-defined, one needs the positivity assumption.

\subsection{On Why the Weights (\ref{eqn:w1_long}) lead to $E(Y_t \mid \bar{a}_t)$}\label{sec:appendix_theory_weights}
Consider the recursive evaluation of the integral (\ref{eqn:estimand_weighted_long}) from $s=t$ to $s=0$, which shows how the weights recover the association:
\begin{align*}
    \tilde{m}_{w,t}(a_t, \mathbf h_t) &= E(Y_t\mid A_{t}=a_t, \mathbf H_t=\mathbf h_t),\\
    \tilde{m}_{w,t-1}(a_{t-1}, \mathbf h_{t-1}) &= \int E(Y_t\mid A_{t}=a_t, \mathbf H_t=\mathbf h_t)d P_0(\mathbf l_{t}\mid A_{t} = a_{t}, A_{t-1} = a_{t-1}, \mathbf H_{t-1}=\mathbf h_{t-1})\\
    &=E[E(Y_t\mid A_{t}, \mathbf H_t)\mid A_{t} = a_{t}, A_{t-1} = a_{t-1}, \mathbf H_{t-1}=\mathbf h_{t-1})]\\
    &=E[Y_t\mid A_{t} = a_{t}, A_{t-1} = a_{t-1}, \mathbf H_{t-1}=\mathbf h_{t-1})]\\
    & \ldots \\
    \tilde{m}_{w,0}(a_0, \mathbf h_{0}) &= E[Y_t\mid A_{t} = a_{t}, A_{t-1} = a_{t-1},\ldots, A_0 = a_0] = E(Y_t \mid \bar{a}_t)
\end{align*}

\subsection{Iterated Nested Expectation Representation of Estimand 2}\label{sec:appendix_theory_estimand2_details}
Using the definition of (\ref{eqn:estimand_general_long}), $\tilde{m}_{t+1} = Y_t$ and observing that $\int \tilde{m}_{t+1}(a_{t+1}, \mathbf h_{t+1})\, w_t(a_{t+1}, \mathbf h_{t+1}) \, d P_0(\mathbf l_{t+1}\mid A_t = a_t, \mathbf H_t=\mathbf h_t)$ = $E(\tilde{m}_{t+1}(a_{t+1}, \mathbf{H}_{t+1})\, w_t(a_{t+1}, \mathbf h_{t+1}) \mid A_t=a_t,\mathbf{H}_t=\mathbf{h}_t)$, we have for $s=t$:
\begin{eqnarray*}
\tilde{m}_t (a_t, \mathbf{h}_t) &=& E(Y_t \, w_t  \mid A_t=a_t,\mathbf{H}_t=\mathbf{h}_t) \,,
\end{eqnarray*}
where $w_t = w_t(a_{t+1}, \mathbf h_{t+1})$. For $s= t-1$, we then get
\begin{eqnarray*}
\tilde{m}_{t-1} (a_{t-1}, \mathbf{h}_{t-1})  &=& E(\tilde{m}_{t} (a_t, \mathbf{h}_t)\,w_{t-1} \mid A_{t-1}=a_{t-1},\mathbf{H}_{t-1}=\mathbf{h}_{t-1}) \\
 &=& E(E(Y_t w_t \mid \bar{A}_t = \bar{a_t}, \bar{L}_t = \bar{l}_t)\, w_{t-1} \mid \bar{A}_{t-1} = \bar{a}_{t-1}, \mathbf{H}_{t-1} = \mathbf{h}_{t-1})\,.
\end{eqnarray*}
Reevaluating the integral from $s=t-2,\ldots,0$ then leads to
\begin{eqnarray*}
\tilde{m}_0 (a_0, \mathbf{h}_0)  &=& \mathbb{E}(\mathbb{E}(\ldots\mathbb{E}(\,\mathbb{E}(Y_t w_t|\bar{{A}}_{t}=\bar{a}_{t}, \mathbf{\bar{L}}_t) \, w_{t-1} | \bar{{A}}_{t-1}=\bar{a}_{t-1}, \mathbf{\bar{L}}_{t-1}\,)\ldots)\, w_0|{A}_{0}={a}_{0}, \mathbf{{L}}_{0}\,)\,)\,.
\end{eqnarray*}

\subsection{Parametric $g$-formula representation}\label{sec:appendix_theory_parametric_gformula}

The weighted estimand, shown in the iterated weighted outcome representation in  (\ref{eqn:seq_g_formula_weighted}) can, alternatively, also be rewritten as
\begin{eqnarray}\label{eqn:gformula_parametric_weighted}
\tilde{m}_{w,t} &=&
\int_{\mathbf{\bar{l}}\in\mathbf{\bar{L}}_t}
\left\{
\begin{aligned}
&\mathbb{E}(w_{t}Y_t|\bar{\mathbf{A}}_{t}=\bar{a}_{t}, \mathbf{\bar{L}}_t=\mathbf{\bar{l}}_t
)\times\\
&\prod_{r=0}^t
  \begin{array}{l}
  f(\mathbf{l}_r|\bar{\mathbf{A}}_{r-1}=\bar{a}_{r-1}, \mathbf{\bar{L}}_{r-1}=\mathbf{\bar{l}}_{r-1})
  \end{array}
\end{aligned}
\right\}
\, d\mathbf{\bar{l}} \,,
\end{eqnarray}
where the second term in (\ref{eqn:gformula_parametric_weighted}) is

\begin{eqnarray*}\label{eqn:gformula_parametric_factor}
&\prod_{s=1}^{q-1}&
f(l_r^s|\bar{\mathbf{A}}_{r-1} = \bar{a}_{r-1}, \mathbf{\bar{L}}_{r-1}=\mathbf{\bar{l}}_{r-1}, L_r^1 = l_r^1,\ldots, L_r^{s-1}=l_r^{s-1})\\
&&\times f(w_{r}Y_r|\bar{\mathbf{A}}_{r}=\bar{a}_{r}, \mathbf{\bar{L}}_r=\mathbf{\bar{l}}_r)\,.
\end{eqnarray*}
Note that in the above notation $\bar{L}_s$ includes past outcomes, multiplied with the respective weights.

The equality follows from the knowledge that iterated nested expectation representations of the g-formula can be reexpressed as a traditional (parametric) g-formula factorization because both representations essentially standardize with respect to the post intervention distribution of the time-dependent confounders \cite{Bang:2005, Schomaker:2019, Petersen:2014a}. In our case, the outcome is $w_{t}Y_t$.

\subsection{Time-to-Event Considerations}\label{sec:appendix_theory_survival}
Suppose we have a time-ordering of $(L_s, A_{s,1}, A_{s,2}, \ldots, A_{s,p}, C_s, Y_s)$, $s=0,\ldots,t$. If we use weights to target $E(Y_t^{\bar{C}_t=0}|A_t=a_t,\ldots,A_0=a_0)$ for units that do not have enough conditional support for the intervention strategy of interest, i.e. $\bar{A}^{(j)}_t$, we have to modify the estimation algorithm of Table \ref{tab:estimation_gcomp_multiple} as follows:\\

{
\small
\renewcommand{\arraystretch}{1.2}
\noindent \begin{tabular}{p{0.0925\textwidth}p{0.875\textwidth}}
\hline
Step 1a/b: & Estimate the densities among those uncensored, and without prior event ($C_t=0, Y_{t-1}=0$).\\
Step 2: & Also, set $c_t=0$.\\
Step 4: &  Fit the model for the weighted outcome among those uncensored and without prior event,\\
&that is, estimate $\mathbb{E}(w_t(a_t^{(1)},c) \, \tilde{Y}_t|\bar{A}_t, C_t=0, Y_{t-1}=0, \mathbf{H}_t)$. \\
Step 5: & Additionally, set $\tilde{Y}_{t-1,i}=1$, if $Y_{t-2,i}=1$.\\
\hline
\end{tabular}
}

\clearpage
\subsection{Example code for the implementation of the algorithm of Table \ref{tab:estimation_gcomp_multiple}}\label{sec:appendix_code_algorithm}
\begin{mdframed}[linecolor=black!50]
\footnotesize{
\begin{verbatim}
# load example efavirenz data from package CICI, select 2 time points for illustration
# variable order: sex metabolic log_age NRTI weight.0 efv.0 VL.0 adherence.1 weight.1 efv.1 VL.1
library(CICI); library(CICIplus); data(EFV); EFV.2 <- EFV[,1:11]

# manual implementation of estimand 2

# Step 0a: interventions are (efv.0, efv.1) = [(0,0),(0.5,0.5),(1,1)]
# Step 0b: define \tilde{Y}_1 = Y_1 = VL.1
Y.tilde.1 <- EFV.2$VL.1

### iteration: t=1
# Step 1a: estimate numerator density g(efv.1 | past)
#          (assuming normality for illustration)
fitted.n.1 <- lm(efv.1 ~ . ,data=subset(EFV.2, select=-VL.1))
# Step 1b: estimate denominator density g(efv.1 | past up to efv.0)
fitted.d.1 <- lm(efv.1 ~ sex+metabolic+log_age+NRTI+weight.0+efv.0,data=EFV.2)
# Step 2: set a_1=a_0=0
EFV.2.A <- EFV.2; EFV.2.A[,c("efv.0","efv.1")] <- 0
# Step 3a: evaluate densities at a=0
eval.n.1 <- dnorm(0,mean=predict(fitted.n.1, newdata=EFV.2.A),sd=summary(fitted.n.1)$sigma)
eval.d.1 <- dnorm(0,mean=predict(fitted.d.1, newdata=EFV.2.A),sd=summary(fitted.d.1)$sigma)
# Step 3b: estimate weights with c=0.01
wf  <- function(num,den,c){as.numeric(num > c) +  as.numeric(num <= c)*(num/den)}
w.1 <- wf(eval.n.1,eval.d.1,c=0.01)
# Step 4: estimate iterated weighted outcome regression
# (binomial as all wY are 0/1 in this specific case, otherwise gaussian)
mY.1 <- glm(I(Y.tilde.1*w.1) ~ ., data=subset(EFV.2, select=-VL.1), family="binomial")
# Step 5: predict iterated weighted outcome under intervention
Y.tilde.0 <- predict(mY.1, newdata=EFV.2.A, type="response")

### iteration: t=0
# Step 1a: estimate numerator density g(efv.0 | past)
fitted.n.0 <- lm(efv.0 ~ sex+metabolic+log_age+NRTI+weight.0 ,data=EFV.2)
# Step 1b: estimate denominator density g(efv.0)
fitted.d.0 <- lm(efv.0 ~ 1 ,data=EFV.2)
# Step 2: set a_0=0 (done above already)
# Step 3a: evaluate densities at a=0
eval.n.0 <- dnorm(0,mean=predict(fitted.n.0, newdata=EFV.2.A),sd=summary(fitted.n.0)$sigma)
eval.d.0 <- dnorm(0,mean=predict(fitted.d.0, newdata=EFV.2.A),sd=summary(fitted.d.0)$sigma)
# Step 3b: estimate weights with c=0.01
w.0 <- wf(eval.n.0,eval.d.0,c=0.01)
# Step 4: estimate iterated (unweighted) outcome regression
mY.0 <- glm(Y.tilde.0 ~ sex+metabolic+log_age+NRTI+weight.0+efv.0, data=EFV.2)
# Step 5: final estimate under intervention
estimate <- weighted.mean(predict(mY.0, newdata=EFV.2.A),w=w.0)

# Step 6: Repeat steps 1-5 for a=(0.5,0.5) and a=(1,1)
# Step 7: Bootstrapping

# Note: the online material contains code on how to arrive at exactly the same result with the package
# Note: typically super learning is used to estimate the iterated outcome regressions, and the
#       densities are preferably estimated non-parametrically. The package offers these options.
\end{verbatim}
}
\end{mdframed}

\clearpage
\section{Additional Results}
\subsection{Simulation and Analysis Results}
\begin{figure}[ht!]
\begin{center}
\subfloat[Simulation 1: weighted curve, $c=1$, $t=2$]{\label{figure:sim_extra1}\includegraphics[scale=0.35]{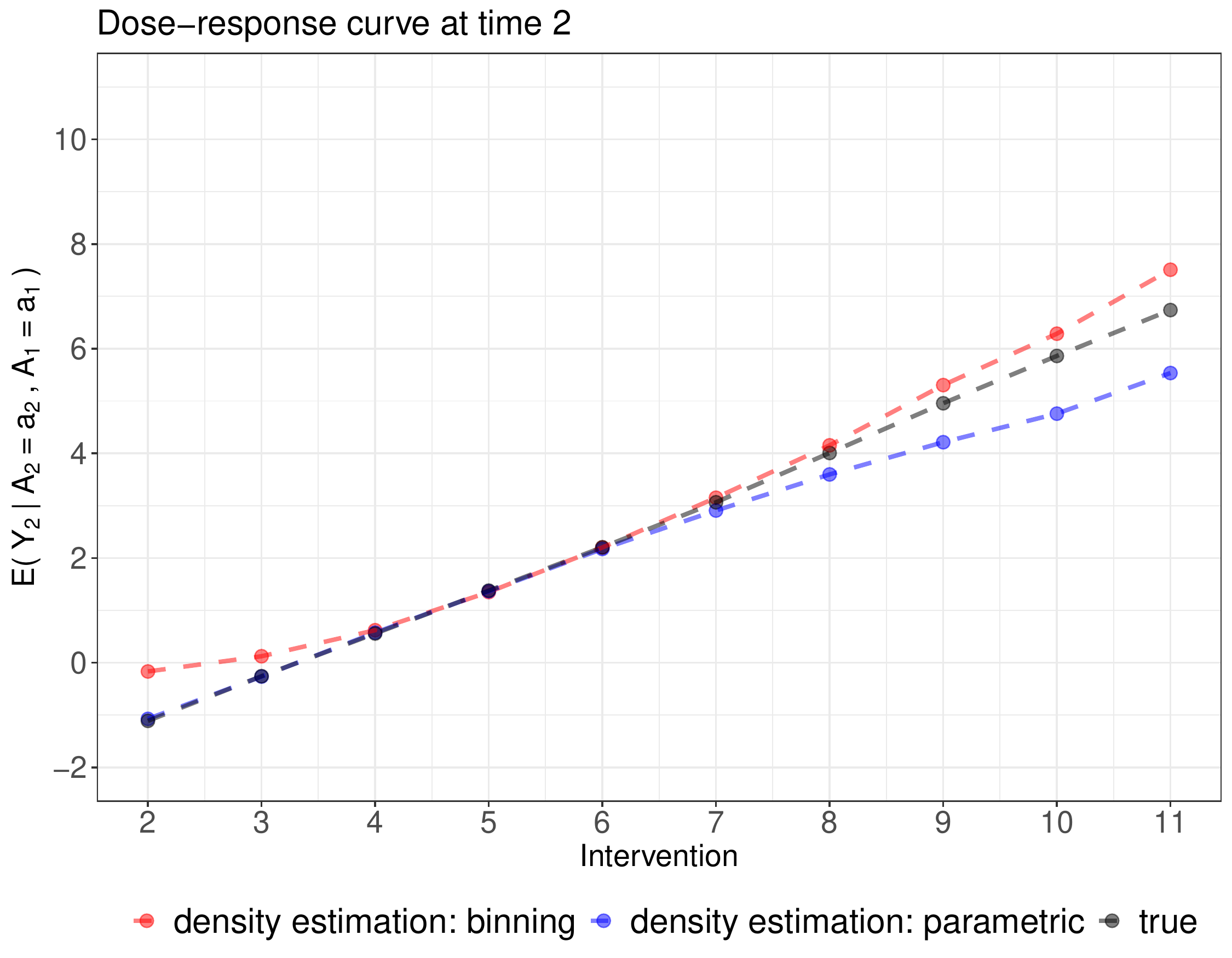}\hfill}
\subfloat[Simulation 2: true and estimated CDRC at $t=2$]{\label{figure:sim_extra2}\includegraphics[scale=0.35]{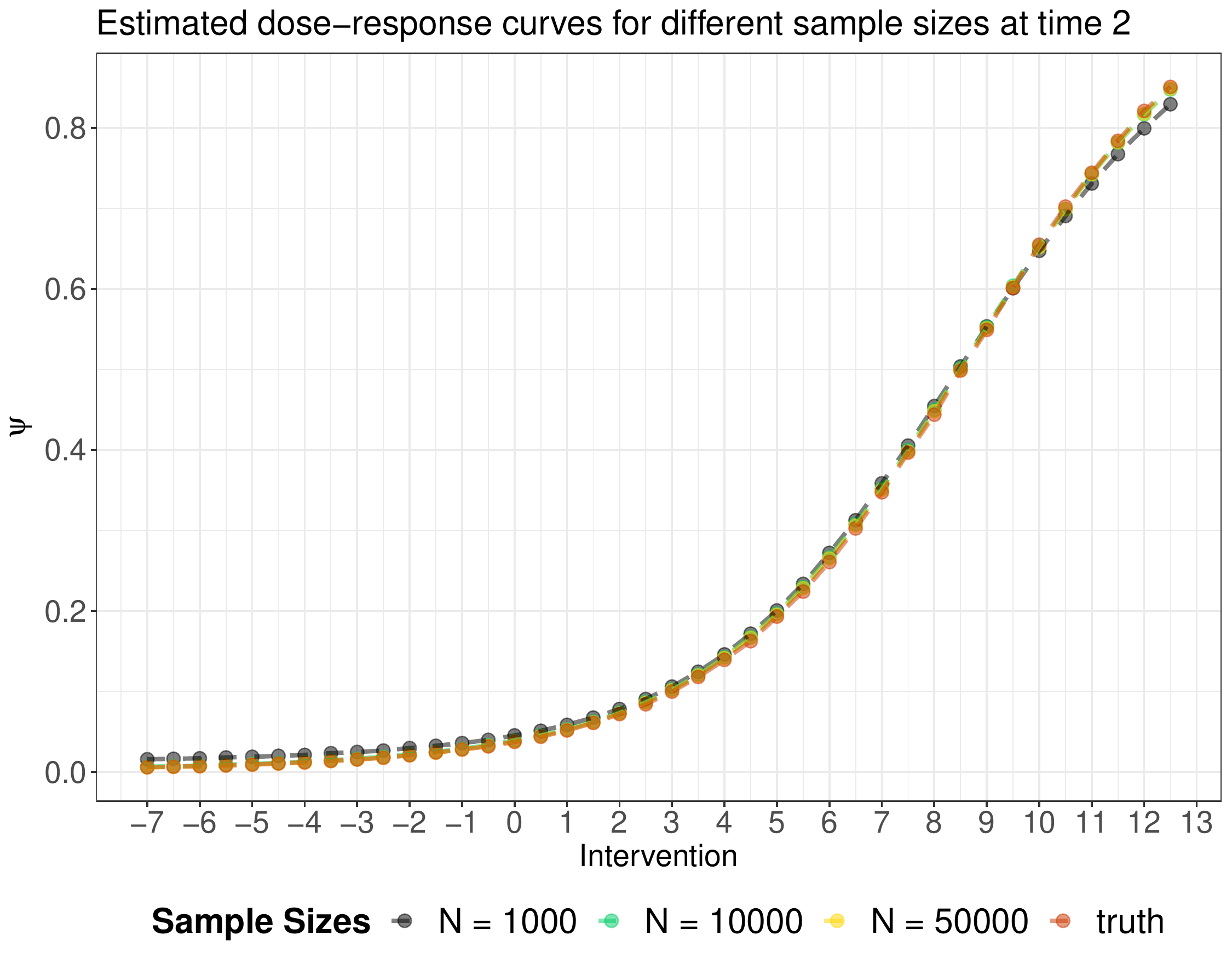}} \\
\subfloat[Simulation 3: true and estimated CDRC at $t=2$]{\label{figure:sim_extra3}\includegraphics[scale=0.35]{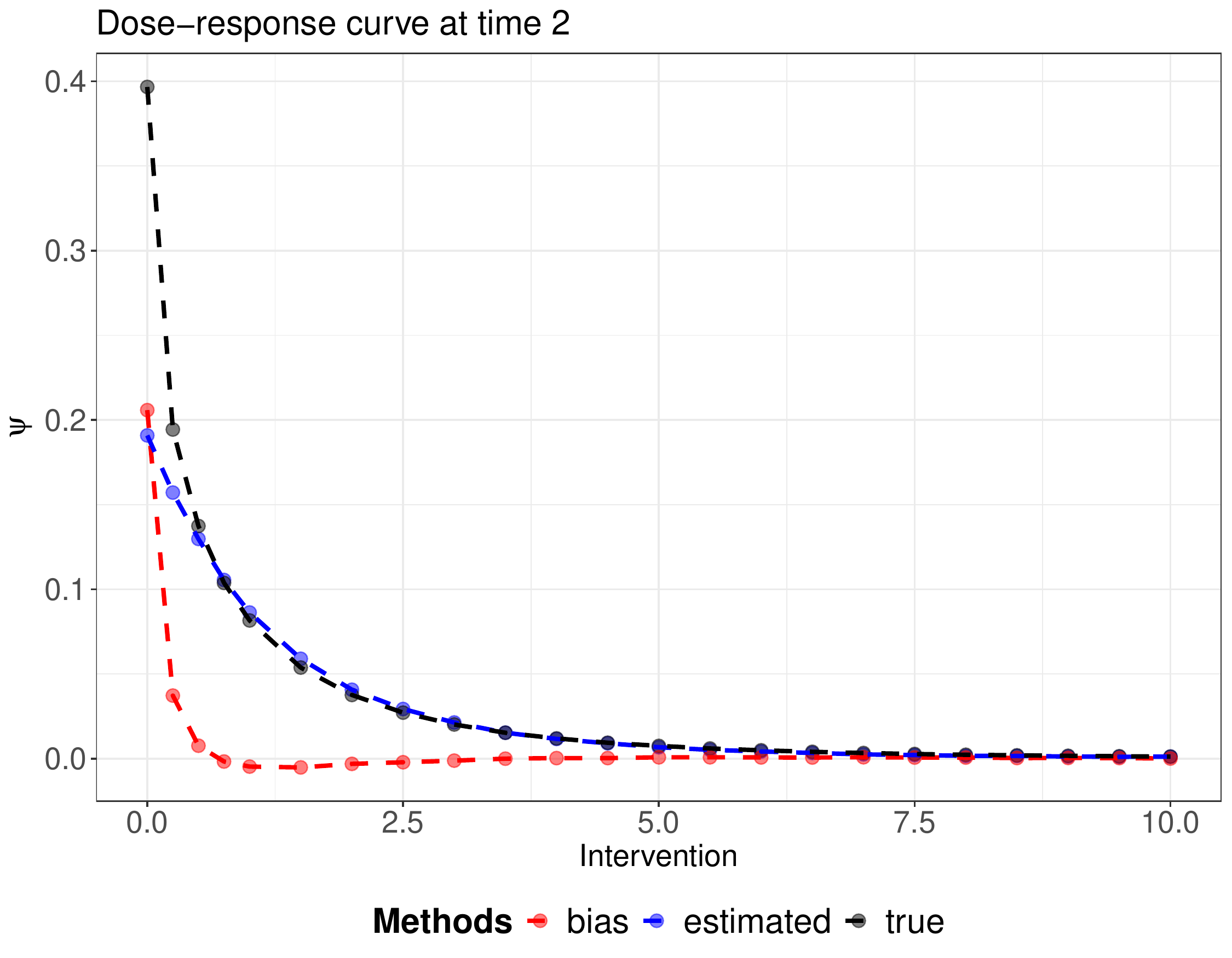}\hfill}
\subfloat[Simulation 3: weighted curve, $c=1$, $t=2$]{\label{figure:sim_extra4}\includegraphics[scale=0.35]{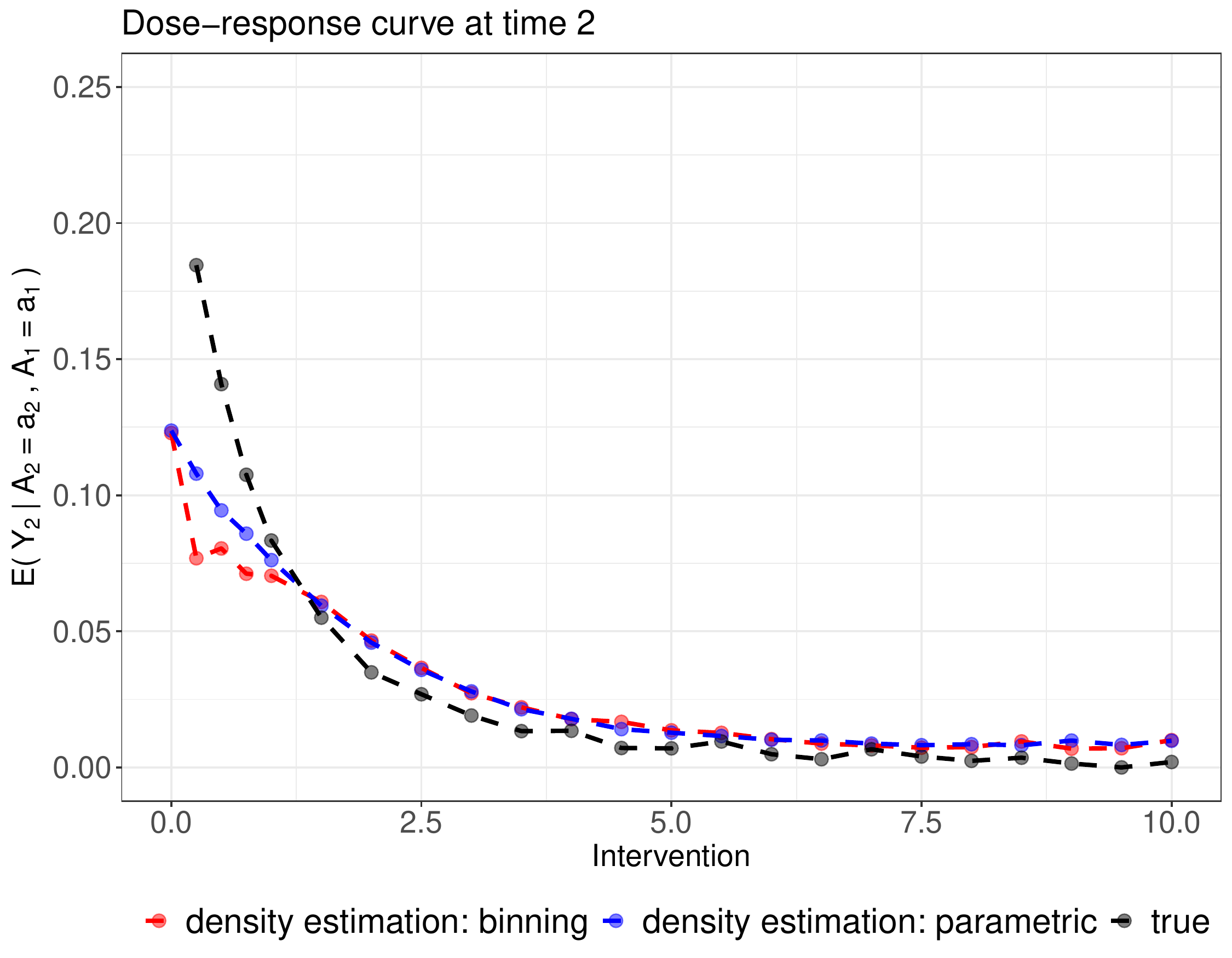}} \\
\subfloat[Simulation 3: CDRC's, $t=2$, $c=0.001, 0.01, 1$]{\label{figure:sim_extra5}\includegraphics[scale=0.35]{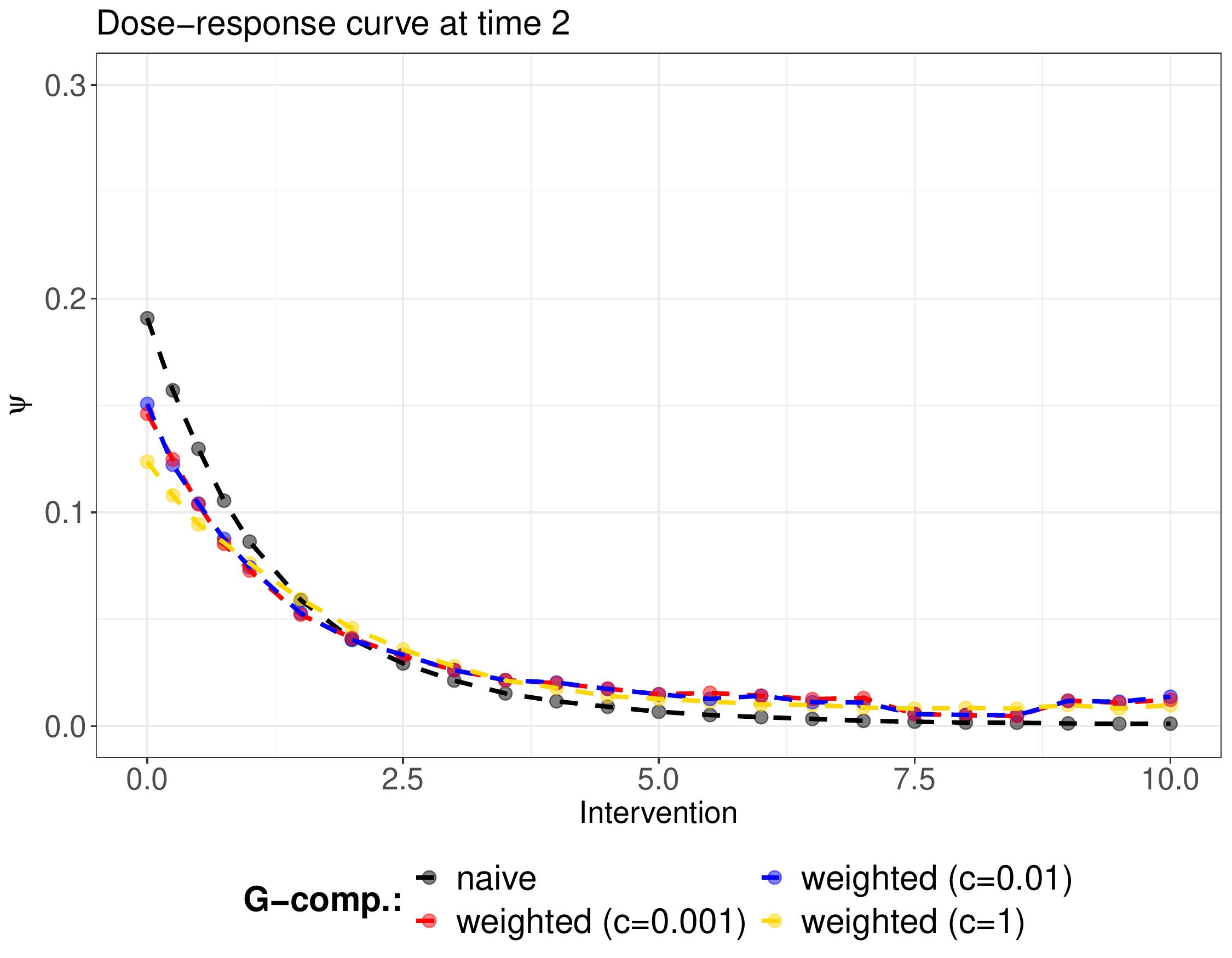}}
\subfloat[Data analysis: estimated CDRC]{\label{figure:data_extra6}\includegraphics[scale=0.35]{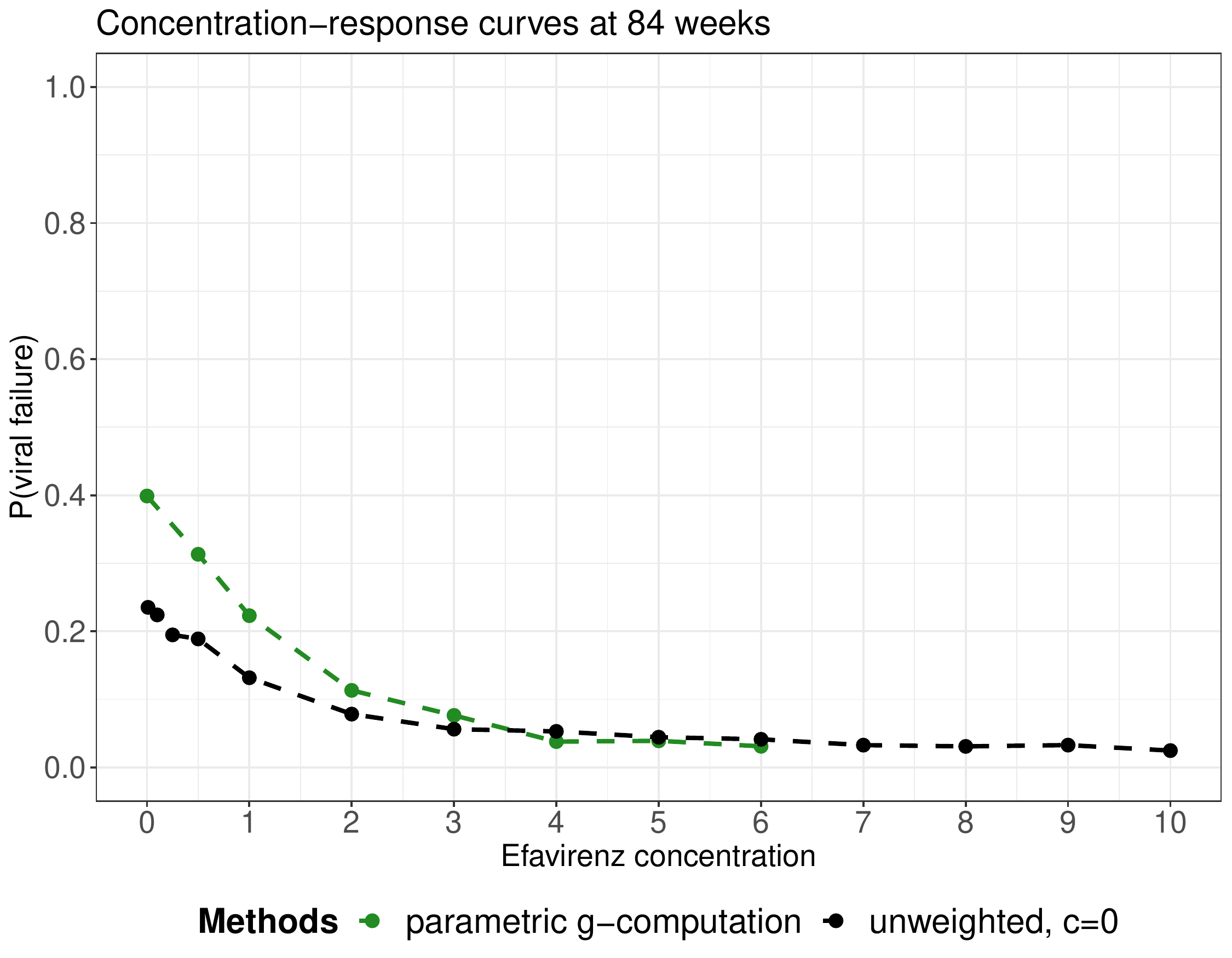}}
\caption{Additional simulation and analysis results}\label{figure:results_extra2}
\end{center}
\end{figure}

\clearpage
\subsection{Intervention Support}
\begin{figure}[ht!]
\begin{center}
\subfloat[Simulation 1]{\label{figure:sim_support1}\includegraphics[scale=0.335]{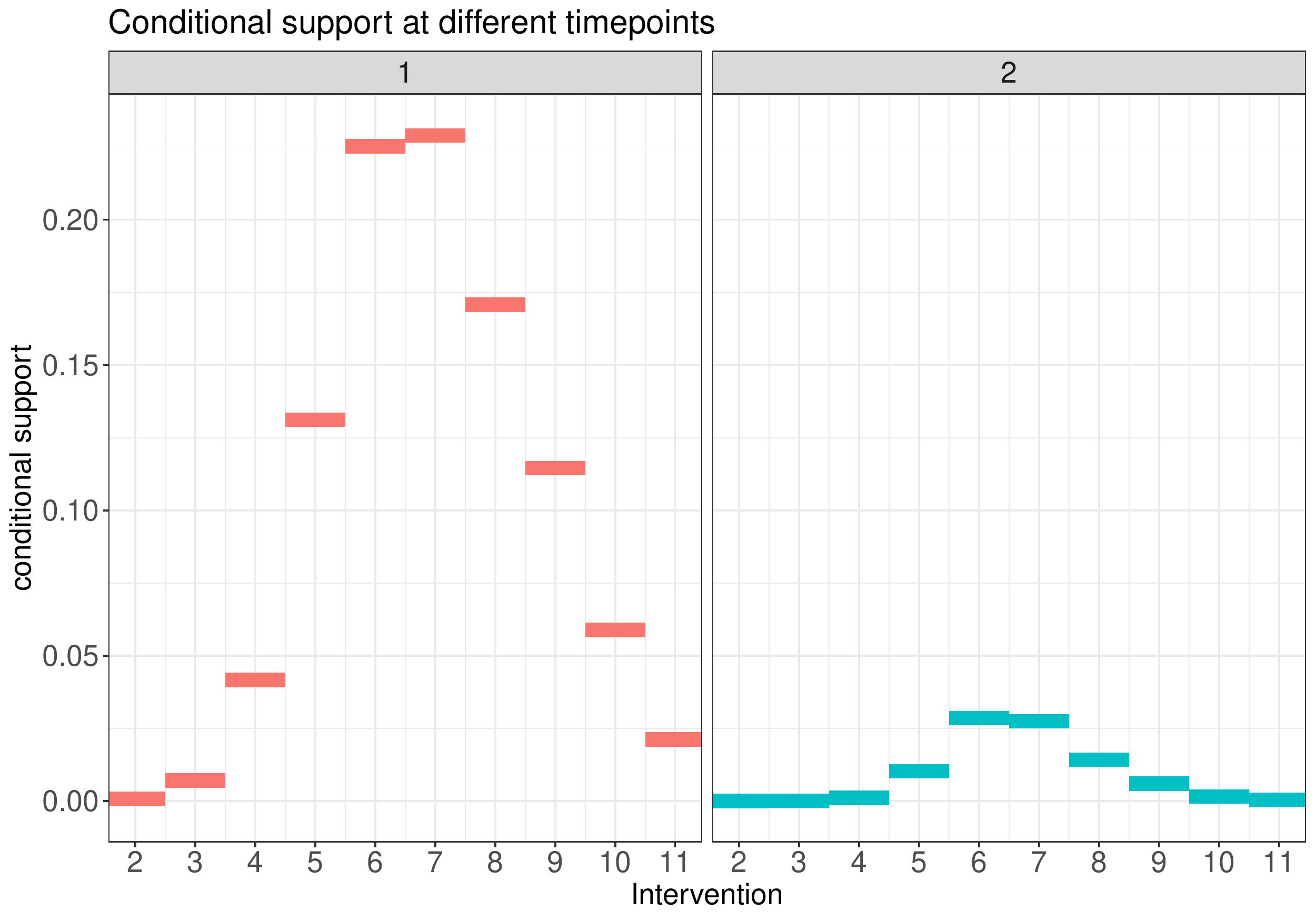}}
\subfloat[Simulation 2]{\label{figure:sim_support2}\includegraphics[scale=0.335]{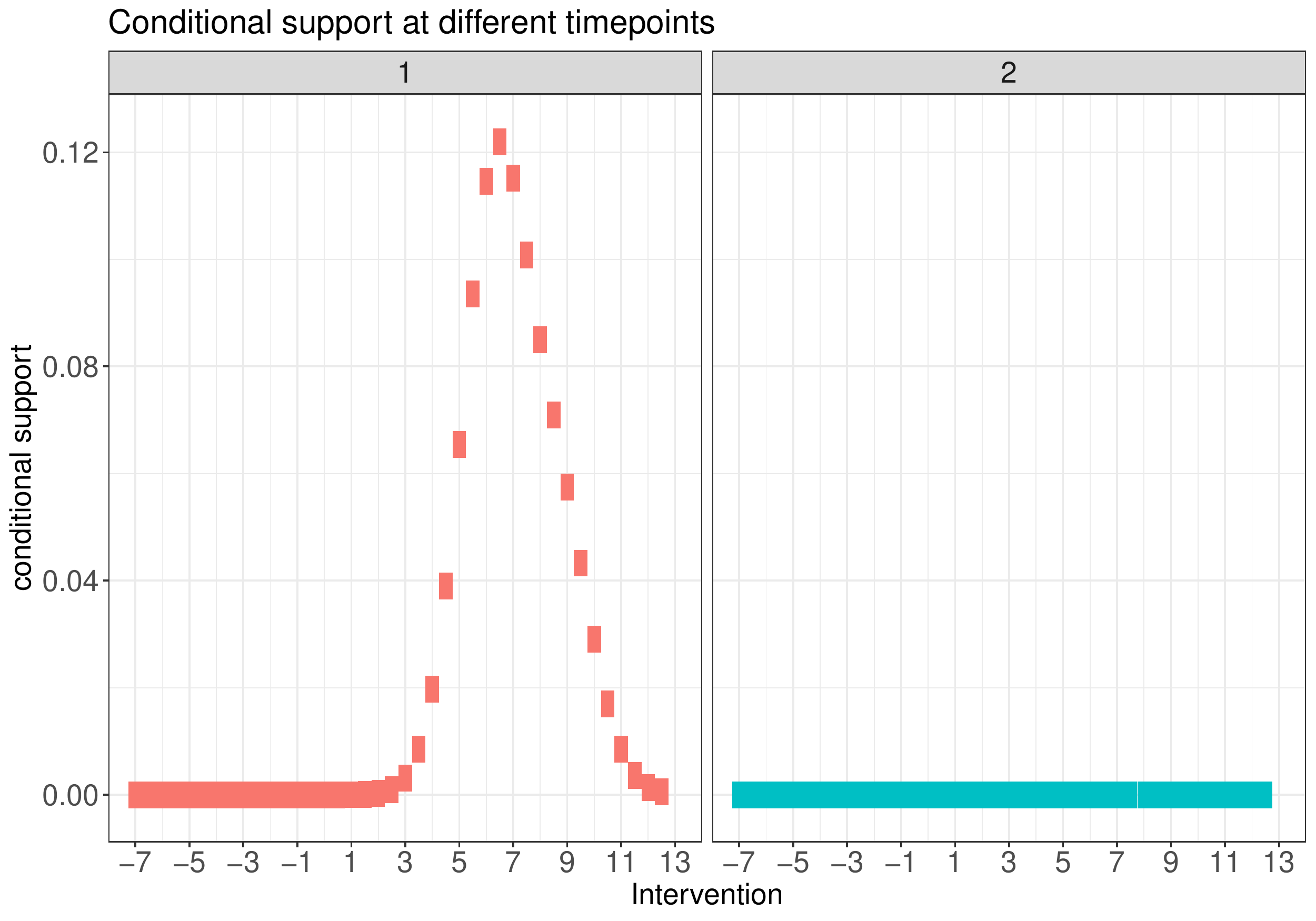}} \\
\subfloat[Simulation 3]{\label{figure:sim_support3}\includegraphics[scale=0.335]{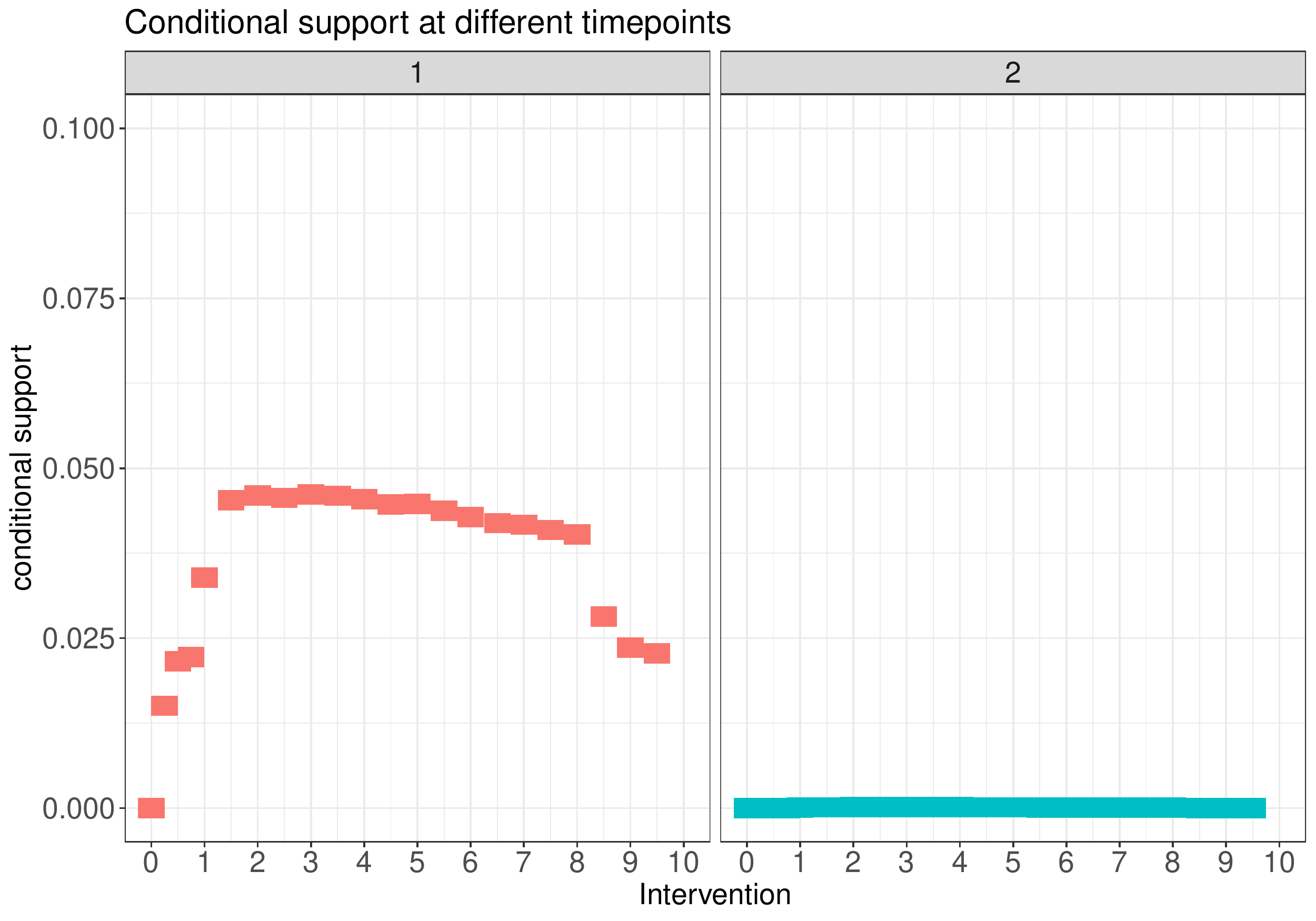}}
\subfloat[Data Analysis]{\label{figure:data_support4}\includegraphics[scale=0.335]{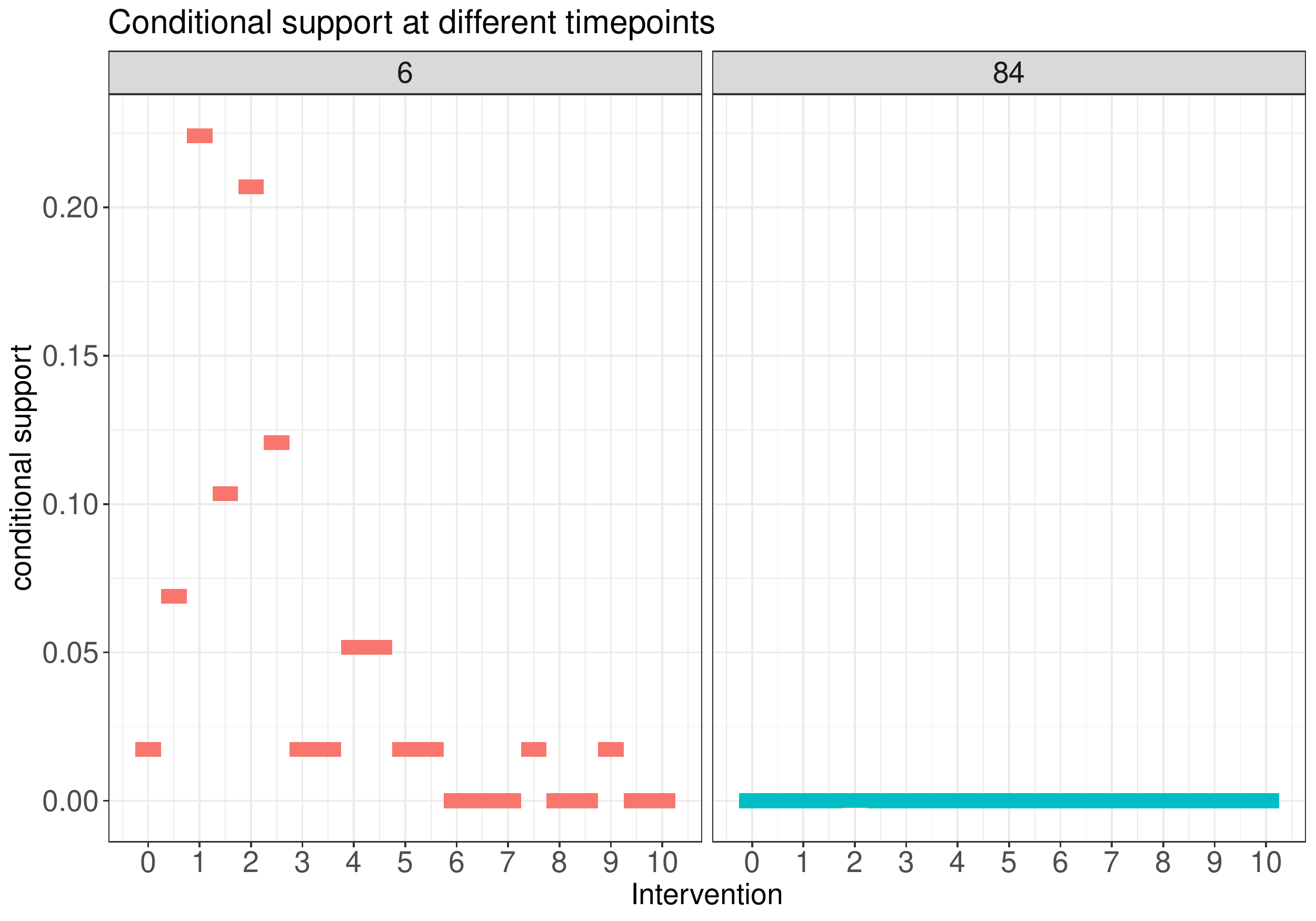}}
\caption{Conditional support for intervention strategies of interest, for both the simulation settings and the data analysis.}\label{figure:results_extra1}
\end{center}
\end{figure}

\section{Data-Generating Processes}\label{sec:appendix_DGP}
\subsection{DGP for Simulation 1}\label{sec:appendix_DGP_1}

{\scriptsize{
\noindent For $t=0$:
\noindent
\begin{eqnarray*}
&& \phantom{\hspace*{15cm}} \\[-0.5cm]
\text{L}^1_0 &\sim& \text{B}(p=0.3) \\
\text{L}^2_0 &\sim& \text{N}(\mu = -1 + 2 \times \text{L}^1_0,\, \sigma=1)\\
\text{A}_0 &\sim& \text{N}(\mu=7+\text{L}^1_0+0.7\times\text{L}^2_0,\,\sigma=1) \\
\text{Y}_0 &\sim& \text{N}(\mu=-1+0.5\times\text{A}_0+0.5\times\text{L}^2_0,\,\sigma=1)
\end{eqnarray*}
\noindent For $t=1,2$:
\noindent
\begin{eqnarray*}
&& \phantom{\hspace*{15cm}} \\[-0.5cm]
\text{L}^1_t &\sim& \text{B}(p=1/(1+\exp(-(-4+\text{L}^1_{t-1}+0.15\times\text{L}^2_{t-1}+0.15\times\text{A}_{t-1})))) \\
\text{L}^2_t &\sim& \text{N}(\mu = 0.5 \times \text{L}^1_t + 0.25\times \text{L}^2_{t-1}+0.5\times\text{A}_{t-1},\, \sigma=1)\\
\text{A}_t &\sim& \text{N}(\mu=\text{A}_{t-1} + \text{L}^1_{t} - 0.1\times\text{L}^2_t,\,\sigma=0.5) \\
\text{Y}_t &\sim& \text{N}(\mu=-2+0.25\times\text{A}_t-0.2\times\text{L}^1_t+\text{L}^2_t,\,\sigma=0.5)
\end{eqnarray*}
}}

\subsection{DGP for Simulation 2}\label{sec:appendix_DGP_2}

{\scriptsize{
\noindent For $t=0$:
\noindent
\begin{eqnarray*}
&& \phantom{\hspace*{15cm}} \\[-0.5cm]
\text{L}^1_0 &\sim& \text{B}(p=0.3) \\
\text{L}^2_0 &\sim& \text{N}(\mu = -1 + 2 \times \text{L}^1_0,\, \sigma=1)\\
\text{A}_0 &\sim& \text{N}(\mu=7+\text{L}^1_0+0.7\times\text{L}^2_0,\,\sigma=1) \\
\text{Y}_0 &\sim& \text{B}(p = 1/(1+\exp(-(-4+0.2\times\text{A}_0+0.5\times\text{L}^2_0))))
\end{eqnarray*}
\noindent For $t=1,2,3,4$:
\noindent
\begin{eqnarray*}
&& \phantom{\hspace*{15cm}} \\[-0.5cm]
\text{L}^1_t &\sim& \text{B}(p=1/(1+\exp(-(-4+\text{L}^1_{t-1}+0.15\times\text{L}^2_{t-1}+0.15\times\text{A}_{t-1})))) \\
\text{L}^2_t &\sim& \text{N}(\mu = 0.5 \times \text{L}^1_t + 0.25\times \text{L}^2_{t-1}+0.5\times\text{A}_{t-1},\, \sigma=1)\\
\text{A}_t &\sim& \text{N}(\mu=-2+0.5\times\text{A}_{t-1} + 0.75\times\text{L}^1_{t} + 0.35\times\text{L}^2_t,\,\sigma=0.5) \\
\text{C}_t &\sim& \text{B}(p=1/(1+\exp(-(-2+0.5\times\text{L}^1_{t}+0.2\times\text{A}_{t}))))  \\
\text{Y}_t &\sim& \text{B}(p = 1/(1+\exp(-(-4+0.2\times\text{A}_t+0.5\times\text{L}^2_t)))) \\
\end{eqnarray*}
}}

\subsection{DGP for Simulation 3}\label{sec:appendix_DGP_3}

Both baseline data ($t=0$) and follow-up data ($t=1,\ldots,4$) were created using structural equations using the $R$-package \texttt{simcausal}. The below listed distributions, listed in temporal order, describe the data-generating process. Our baseline data consists of sex, genotype, $\log(\text{age})$, $\log(\text{weight})$ and the respective Nucleoside Reverse Transcriptase Inhibitor (NRTI). Time-varying variables are co-morbidities (CM), dose, efavirenz mid-dose concentration (EFV), elevated viral load (= viral failure, VL) and adherence (measured through memory caps, MEMS), respectively.
In addition to Bernoulli ($B$), Multinominal ($MN$) and Normal ($N$) distributions, we also use truncated normal distributions;  they are denoted by $N_{[a,a_1,a_2,b,b_1,b_2]}$, where $a$ and $b$ are the truncation levels. Values which are smaller than $a$ are replaced by a random draw from a $U(a_1,a_2)$ distribution and values greater than $b$ are drawn from a $U(b_1,b_2)$ distribution, where $U$ refers to a continuous uniform distribution. For the specified multinomial distributions, probabilities are normalized, if required, such that they add up to 1. The data-generating process reflects the following considerations: more complexity than the first two simulation settings in terms of distribution shape and variety, as well as non-linearities; similarity to the assumed DGP in the data analysis; generation of both areas of poor and good conditional intervention support, such that the proposed weighting scheme can be evaluated in all its breadth.
\vspace*{0.5cm}

{\scriptsize{
\noindent For $t=0$:
\noindent
\begin{eqnarray*}
\text{Sex}_0 &\sim& \text{B}(p=0.5) \\
\text{Genotype}_0 &\sim& \text{MN} \left( \begin{array}{ll}
               p1= 1/( 1+ \exp(-(-0.103 + \text{I}(\text{Sex}_0 = 1) \times 0.223 + \text{I}(\text{Sex}_0 = 0) \times 0.173)))\,,\\
               p2= 1/( 1+ \exp(-(-0.086 +  \text{I}(\text{Sex}_0 = 1) \times 0.198 + \text{I}(\text{Sex}_0 = 0) \times 0.214)))\,,\\
               p3= 1/( 1+ \exp(-(-0.090+  \text{I}(\text{Sex}_0 = 1) \times 0.082 +  \text{I}(\text{Sex}_0 = 0) \times  1.070 )))\\
               \end{array}
               \right) \\
\text{Age}_0 &\sim & N_{[0.693, 0.693, 1, 2.8, 2.7, 2.8]}( \mu = 1.501, \sigma = 0.369) \\
\text{Weight}_0 &\sim & N_{[2.26, 2.26, 2.67, 3.37, 3.02, 3.37]}( \mu = (1.5 + 0.2 \times \text{Sex} + 0.774 \times \text{Age}) \times 0.94), \sigma = 0.369) \\
\text{NRTI}_0 &\sim& \text{MN} \left(  \begin{array}{ll}
               p1= 1/( 1+ \exp(-(-0.006 + \text{I}(\text{Age}_0>1.4563) \times  \text{Age}_0 \times 0.1735 +
               \text{I}(\text{Age}_0 \leq 1.4563) \times \text{Age}_0 \times 0.1570)))\,,\\
               p2= 1/( 1+ \exp(-(-0.006  + \text{I}(\text{Age}_0 > 1.4563) \times  \text{Age}_0  \times 0.1735 +
               \text{I}(\text{Age}_0 \leq 1.4563) \times \text{Age}_0 \times 0.1570)))\,,\\
               p3= 1/( 1+ \exp(-(-0.006  + \text{I}(\text{Age}_0 > 1.4563) \times  \text{Age}_0  \times 0.1570 +
               \text{I}(\text{Age}_0 \leq .14563) \times \text{Age}_0 \times 0.1818))) \\
               \end{array}
               \right) \\
\text{CM}_0 & \sim & \text{B}(p=0.15)\\
\text{Dose}_0 & \sim &\text{MN} \left( \begin{array}{ll}
              p1= 1/( 1+ \exp(-(5 + \sqrt{(\text{Weight}_0)} \times 8 -  \text{Age}_0 \times 10)))\,, \\
              p2= 1/( 1+ \exp(-(4 + \sqrt{(\text{Weight}_0)} \times 8.768 - \text{Age}_0 \times 9.06))) \,, \\
              p3= 1/( 1+ \exp(-(3 + \sqrt{(\text{Weight}_0)} \times 6.562 - \text{Age}_0 \times 8.325)))\,,\\
              p4= 1-(p1+p2+p3) \\
               \end{array}
               \right) \\
\text{EFV}_0 &\sim & N_{[0.2032, 0.2032, 0.88, 21, 8.376, 21]}(\mu = -8 + \text{Age}_0 \times 0.1 + \text{Genotype}_0 \times 4.66 + \text{Dose}_0 \times 0.1 + \\  && \text{I}(\text{Genotype}_0 \leq 2) \times 2.66 +  \text{I}(\text{Genotype}_0 = 3) \times 4.6,  \sigma = 4.06)\\
\text{VL}_0 & \sim & \text{B}(p= 1 - (1/(1+\exp(-(0.4+1.9 \times \sqrt{\text{EFV}_0})))))\\
\end{eqnarray*}

\noindent For $t\geq1$:
\noindent
\begin{eqnarray*}
\text{MEMS}_t &\sim& \text{B}(p= 1/(1+ \exp(-(0.71 + \text{CM}_{t-1} \times 0.31 + \text{MEMS}_{t-1} \times \text{I}(t\geq2) \times 0.31)))) \\
&& \text{with}\,\, \text{MEMS}_{t-1} \text{used as of} \,\, t=2\\
\text{Weight}_t &\sim& N_{[2.26,2.26,2.473,3.37,3.2,3.37]}(\mu = \text{Weight}_{t-1}\times 1.04 -0.05 \times \text{I}(\text{CM}_{t-1}=1),\, \sigma=0.4 )  \\
\text{CM}_t &\sim & \text{B}(p = 1 - (1/( 1+ \exp(- (0.5 \times \text{I}(\text{CM}_{t-1}=1) + \text{Age}_{0} \times 0.1 + \text{Weight}_{t-1} \times 0.1))))) \\
\text{Dose}_t & \sim & \text{MN} \left(\begin{array}{ll}
               p1 = (1/( 1+ \exp(-(4 + \text{Dose}_{t-1} \times 0.5 + \sqrt{\text{Weight}_t} \times 4  -  \text{Age}_0 \times 10)))\,,\\
               p2 = (1/( 1+ \exp(-(-8 + \text{Dose}_{t-1} \times 0.5  + \sqrt{\text{Weight}_t} \times 8.568  -  \text{Age}_0 \times 9.06)))\,,\\
               p3 = (1/( 1+ \exp(-(20 + \text{Dose}_{t-1} \times 0.5 + \sqrt{\text{Weight}_t} \times 6.562  -  \text{Age}_0 \times 18.325)))\,,\\
               p4 = 1-(p1+p2+p3) \\
               \end{array}
               \right) \\
\text{EFV}_t &\sim & N_{[0.2032, 0.2032, 0.88, 21.84, 8.37, 21.84]}(\mu = 0.1 \times \text{DOSE}_t + 0.1 \times \text{MEMS}_{t} + \text{I}(\text{Genotype}_0 \leq 2) \times 2.66 \\&& \quad +  \text{I}(\text{Genotype}_0 = 3) \times 4.6,  \sigma = 4.06)\\
\text{VL}_t & \sim & \text{B}(p= 1- (1/(1+\exp(-(1-0.6\times\text{I}(t=1)-1.2\times\text{I}(t=4) + 0.1 \times \text{CM}_{t-1} + (2 - 0.2\times\text{I}(t=3)) \times \sqrt{\text{EFV}_t})))))
\end{eqnarray*}
}}

\section{More on the DAG}\label{sec:appendix_DAG}
The measured efavirenz concentration depends on the following factors: the dose itself (which is recommended to be assigned based on the weight-bands), adherence (without regular drug intake, the concentration becomes lower), and the metabolism characterized in the gene CYP2B6, through the 516G and 983T polymorphisms \cite{Bienczak:2016b}. Given the short half-life of the drug relative to the measurement interval, no arrow from EFV$_t$ to EFV$_{t+1}$ is required. Viral failure is essentially caused if there is not enough drug concentration in the body; and there might be interactions with co-morbidites and co-medications. Note also that co-morbidities, which are reflected in the DAG, are less frequent in the given data analysis as trial inclusion criteria did not allow children with active infections, treated for tuberculosis and laboratory abnormalities to be enrolled into the study. Both weight and MEMS (=adherence) are assumed to be time-dependent confounders affected by prior treatments (=concentrations). Weight affects the concentration indirectly through the dosing, whereas adherence affects it directly. Adherence itself is affected by prior concentrations, as too high concentration values can cause nightmares and other central nervous system side effects, or strong discomfort, that might affect adherence patterns. Weight is affected from prior concentration trajectories through the pathway of viral load and co-morbidites. Finally, both weight and adherence affect viral outcomes not only through EFV concentrations, but potentially also through co-morbidities such as malnutrition, pneumonia and others.

\end{document}